\documentstyle[preprint,amstex,aps,eqsecnum,epsfig]{revtex}
\tighten
\def\E{{\cal E}}                            % F
\def\N{{\cal N}}                            % N
\def\T{{\cal T}}                            % T 
\def\V{{\cal V}}                            % V 
\def\a{\alpha}

\def\L{\Lambda}

\def\l{\lambda}

\def\s{\sigma}
\def\om{\omega}

\def\t{\theta}

\def\ZZ{{\Bbb Z }}
\def\bds#1{\boldsymbol{#1}}
\def\lowmp{\lower.11em\hbox{${\scriptstyle\mp}$}}
\def\intf{\int_{-\infty}^{+\infty}}

\def\VEV#1{\left\langle #1 \right\rangle}   % Vacuum Expect. Value
\def\bra#1{\left\langle #1\right|}               % bra
\def\ket#1{\left| #1\right\rangle}               % ket
\def\der#1#2{{d#1\over d#2}}
\def\pdif#1#2{{\partial #1 \over \partial #2}}
\def\braket#1#2{\left\langle #1 \right| \left. #2 \right\rangle}
\def\VEV#1{\left\langle #1 \right\rangle}
\def\vev#1{\langle #1 \rangle}
\def\mbare{m^2_{\text{b}}}
\def\lbare{\lambda_{\text{b}}}

\begin{document}
\preprint{ IFUM 646/FT-99 \quad Bicocca-FT-99-18\bigskip}
%\draft
\title{\bf Out--of--equilibrium dynamics of $\phi^4$ QFT in finite volume}
\author{{\bf C. Destri $^{(a,b)}$}  and {\bf E. Manfredini $^{(b)}$\bigskip}}

\bigskip

\address{
(a) Dipartimento di Fisica G. Occhialini, \\
   Universit\`a di Milano--Bicocca  and INFN, sezione di Milano$^{ 1,2}$
    \\
(b)  Dipartimento di Fisica,  Universit\`a di Milano \\ 
     and INFN, sezione di Milano$^{ 1,2}$
}
\footnotetext{
$^1$mail address: Dipartimento di Fisica, Via Celoria 16, 20133 Milano,
ITALIA.}
\footnotetext{
$^2$e-mail: claudio.destri@@mi.infn.it, emanuele.manfredini@@mi.infn.it
}
\date{June 1999}
\maketitle
\begin{abstract}
The $\lambda \phi^4$ model in a finite volume is studied in the
infinite $N$ limit and within a non--gaussian Hartree--Fock
approximation both at equilibrium and out of equilibrium, with
particular attention to certain fundamental features of the broken
symmetry phase. The numerical solution of the dynamical evolution
equations show that the zero--mode quantum fluctuations cannot grow
macroscopically large starting from microscopic initial
conditions. Thus we conclude that there is no evidence for a dynamical
Bose--Einstein condensation. On the other hand, out of equilibrium the
long--wavelength fluctuations do scale with the linear size of the
system, signalling dynamical infrared properties quite different from
the equilibrium ones characteristic of the same approximation
schemes. This result suggests the cause, and the possible remedy, of
some unlikely features of the application to out--of--equilibrium
dynamics of the standard HF factorization scheme,
which coincides with the gaussian restriction of our Hartree--Fock
approximation.
\end{abstract}

\newpage

\section{Introduction}\label{int}
In the last few years a great deal of attention has been paid to the
study  of interacting quantum fields out of equilibrium. There are,
in fact, many interesting physical situations in which the
standard S--matrix approach cannot give sensible information about
the behavior of the system, because it evolves through a series of 
highly excited states (i.e., states of finite energy density). 

As an example consider any model of cosmological inflation:
it is not possible to extract precise predictions on physical
observables without including in the treatment the quantum
back--reaction of the field on the space--time geometry and on itself
\cite{devega7,kls,riotto}.

On the side of particle physics, the ultra-relativistic heavy-ion
collisions, scheduled in the forthcoming years at CERN--SPS, BNL--RHIC
and CERN--LHC, are supposed to produce hadron matter at very high
densities and temperatures; in such a regime the usual approach based
on particle scattering cannot be considered a good interpretative tool
at all. To extract sensible information from the theory new
computational schemes are necessary, that go beyond the simple
Feynmann diagram expansion. The use of resummation schemes, like the
Hartree--Fock (HF) \cite{tdHF,cactus} approximation and the large $N$
limit (LN) \cite{largen_exp}, or the Hard Thermal Loop resummation for
systems at finite temperature (HTL) \cite{HTL}, can be considered a
first step in this direction. They, in fact, enforce a sum over an
infinite subset of Feynmann diagrams that are dominant in a given
region of the parameter space, where the simple truncation of the
usual perturbative series at finite order cannot give sensible answers.

Quite recently HF and LN have been used in order to clarify some
dynamical aspects of the large $N$ $\phi^4$ theory, reaching the
conclusion that the non--perturbative and non--linear evolution of the
system might eventually produce the onset of a form of non--equilibrium
Bose--Einstein condensation of the long--wavelength Goldstone bosons
usually present in the broken symmetry phase
\cite{bdvhs,relax,tsunami} of the model. Another very interesting
result in \cite{bdvhs} concerns the dynamical Maxwell construction,
which reproduces the flat region of the effective potential in case of
broken symmetry as asymptotic fixed points of the background evolution.  

In this article we present a detailed study, {\em in finite volume},
of dynamical evolution out of equilibrium for the $\Phi^4$ scalar
field. More precisely, we determine how such dynamics scales with the
size of the periodic box containing the system in the case of uniform
backgrounds. This is necessary to address questions like
out--of--equilibrium symmetry breaking and dynamical Bose--Einstein
condensation.
 
We apply two of the non--perturbative methods mentioned above, namely
the Hartree--Fock approximation and the large $N$ expansion.

In section \ref{cft} we define the model in finite volume, giving all
the relevant notations and definitions. We also stress the
convexity of the effective potential as an exact result, valid for the
full renormalized theory in any volume.

In section \ref{ln} we derive the large $N$ approximation of the
$O(N)-$invariant version of $\lambda (\bds\phi ^2)^2$ model, according
to the general rules of ref. \cite{yaffe}. In this derivation it
appears evident the essential property of the $N\to\infty$ limit of
being a particular type of {\em classical} limit, so that it leads to
a classical phase space, a classical hamiltonian with associated
Hamilton's equations of motion [see eqs. (\ref{sk2}), (\ref{Nunbgap})
and (\ref{Nbgap})]. We then minimize the hamiltonian function(al) and
determine the conditions when massless Goldstone bosons
(i.e. transverse fluctuations of the field) to form a Bose--Einstein
condensate, delocalizing the vacuum field expectation value (cfr. also
ref. \cite{chkm}). This necessarily requires that the width of the
zero--mode fluctuations becomes macroscopically large, that is of the
order of the volume. Only when the background takes one of the extremal
values proper of symmetry breaking the width of the
zero--mode fluctuations is of order $L^{1/2}$, as typical of a free
massless spectrum.

The study of the lowest energy states of the model is needed for
comparison with the results of the numerical simulations, which show
that the zero--mode width $\s_0$ stays microscopic (that is such that
$\s_0/$volume$\to 0$ when the volume diverges) whenever it starts from
initial conditions in which it is microscopic. Our results, in fact,
show clearly the presence of a time scale $\tau_L$, proportional to
the linear size $L$ of the system, at which finite volume effects
start to manifest. The most remarkable consequence of the presence of
such a scale is that it prevents the zero mode amplitude to grow
macroscopically large.  This result contradicts the interpretation of
the linear late--time growth of the zero--mode width as a
Bose--Einstein condensation of Goldstone bosons
\cite{bdvhs,relax,tsunami}.

On the other hand we do find that the size of the low--lying widths at
time $\tau_L$ is of order $L$, to be compared to the equilibrium
situation where they would be of order $L^0$ in the massive case or of
order $L^{1/2}$ in the massless case. Perhaps the denomination
``microscopic'' should be reserved to this two
possibilities. Therefore, since our initial condition are indeed
microscopic in this restricted sense, we do observe in the
out--of--equilibrium evolution a rapid transition to a different
regime intermediate between the microscopic one and the macroscopic
one characteristic of Bose--Einstein condensation.

At any rate, when one considers microscopic initial conditions
for the choice of bare mass which corresponds to broken symmetry, the
role itself of symmetry breaking is not very clear in the large $N$
description of the out--of--equilibrium dynamics, making equally
obscure the issues concerning the so--called quantum phase ordering
\cite{bdvhs}. This is because the limit $N\to\infty$ is completely
saturated by gaussian states, which might signal the onset of symmetry
breaking only developing macroscopically large fluctuations. Since
such fluctuations do not appear to be there, the meaning itself of
symmetry breaking, as something happening as times goes on and
accompanied by some kind of phase ordering, is quite unclear.  Of
course, in this respect the main limitation of our approach, as well
as of those of the references mentioned above, is in the assumption of
a uniform background. Nonetheless, phenomena like the asymptotic
vanishing of the effective mass and the dynamical Maxwell
construction, taking place in this contest of a uniform background and
large $N$ expansion, are certainly very significant manifestations of
symmetry breaking and in particular of the Goldstone theorem which
applies when a continuous symmetry is broken.

To gather more information on these matters, we consider in section
\ref{tdhf} a more elaborate type of time--dependent Hartree--Fock (tdHF)
approximation, which generalizes the standard gaussian self-consistent
approach to non--gaussian wave--functionals; in fact, one might
envisage the possibility that, while gaussian fluctuations never
become of the size of the volume, non--gaussian fluctuations do grow in
time to a macroscopic size. We derive therefore the mean--field
coupled time--dependent Schroedinger equations for the modes of the
scalar field, under the assumption of a uniform condensate, see eqs
(\ref{Schroedinger}), (\ref{H_k}) and (\ref{omvev}). A
significant difference with respect to previous tdHF approaches
\cite{devega2} concerns the renormalization of ultraviolet
divergences. In fact, by means of a single proper substitution of the
bare coupling constant $\lbare$ with the renormalized one $\l$ in the
Hartree--Fock hamiltonian, we obtain completely cut-off independent
equations (apart from the corrections in inverse power of the cutoff
which are there due to the Landau pole). The substitution is
introduced by hand, but is justified by simple diagrammatic
considerations.

One advantage of not restricting a priori the self-consistent HF
approximation to gaussian wave--functionals, is in the possibility of
a better description of the structure of the vacuum in case of broken
symmetry. In fact we can show quite explicitly that, in any finite
volume, the ground state the zero--mode of the $\phi$ field is
concentrated around the two vacua of the broken symmetry, driving the
probability distribution for any sufficiently wide smearing of the
field into a two peaks shape. This is indeed what one would
intuitively expect in case of symmetry breaking. On the other hand
none of this appears in a dynamical evolution that starts from a
distribution localized around a single value of the field in the
spinodal region, confirming what already seen in the large $N$
approach. More precisely, within a further controlled gaussian
approximation of our tdHF approach, one observe that initially
microscopic quantum fluctuations never becomes macroscopic, suggesting
that also non--gaussian fluctuations cannot reach macroscopic sizes.
As a simple confirmation of this fact, consider the completely
symmetric initial conditions $\vev{\phi}=\vev{\dot\phi}=0$ for the
background: in this case we find that the dynamical equations for
initially gaussian field fluctuations are identical to those of large
$N$ [apart for a rescaling of the coupling constant by a factor of
three; compare eqs. (\ref{sk2}) and (\ref{Emotion})], so that we
observe the same asymptotic vanishing of the effective mass. However,
this time no interpretation in terms of Goldstone theorem is possible,
since the broken symmetry is discrete; rather, if the width of the
zero--mode were allowed to evolve into a macroscopic size, then the
effective mass would tend to a positive value, since the mass in case
of discrete symmetry breaking is indeed larger than zero.

On the other hand, also in the gaussian HF approach, we do find that
the size of the low--lying widths at time $\tau_L$ is of order $L$.
We then discuss why this undermine the self--consistency of the 
gaussian approximation, imposing the need of further study, both
analytical and numerical.

Finally, in section \ref{conclusion} we summarize the results
presented in this article and we sketch some interesting open problems
that we plan to study in forthcoming works.

\section{Cutoff field theory}\label{cft}

We consider the scalar field operator $\phi$ in a $D-$dimensional
periodic box of size $L$ and write its Fourier expansion as customary
\begin{equation*}
	\phi(x)=L^{-D/2}\sum_k \phi_k\,e^{ik\cdot x} 
	\;,\quad \phi_k^\dag = \phi_{-k}
\end{equation*}
with the wavevectors $k$ naturally quantized: $k=(2\pi/L)n$, $n\in\ZZ^D$.
The canonically conjugated momentum $\pi$ has a similar expansion
\begin{equation*}
	\pi(x)=L^{-D/2}\sum_k \pi_k\,e^{ik\cdot x} 
	\;,\quad \pi_k^\dag = \pi_{-k}
\end{equation*}
with the commutation rules $[\phi_k\,,\pi_{-k'}]=i\delta^{(D)}_{kk'}$.
The introduction of a finite volume should be regarded as a
regularization of the infrared properties of the model, which allows
to ``count'' the different field modes and is needed especially in the
case of broken symmetry.

To regularize also the ultraviolet behavior, we restrict the sums
over wavevectors to the points lying within the $D-$dimensional sphere
of radius $\Lambda$, that is $k^2\le\Lambda^2$, with $\N=\Lambda
L/2\pi$ some large integer. Clearly we have reduced the original
field--theoretical problem to a quantum--mechanical framework with
finitely many (of order $\N^{D-1}$) degrees of freedom.

The $\phi^4$ Hamiltonian reads
\begin{equation*}
\begin{split}
  H &=\dfrac12\int d^Dx\,\left[\pi^2 + (\partial\phi)^2 + \mbare\,\phi^2
	+ \lbare\, \phi^4 \right] = \\
    &=\dfrac12\sum_k\left[\pi_k\pi_{-k} +(k^2+ \mbare)\,\phi_k\phi_{-k}\right]
	+\dfrac\l4 \sum_{k_1,k_2,k_3,k_4}\phi_{k_1}\phi_{k_2}
	 \phi_{k_3}\phi_{k_4}\,\delta^{(D)}_{k_1+k_2+k_3+k_4,0}
\end{split}
\end{equation*}
where $\mbare$ and $\lbare$ should depend on the UV cutoff $\Lambda$
in such a way to guarantee a finite limit $\Lambda\to\infty$ for all
observable quantities. As is known
\cite{wilson,devega2%,froh,latt,broken
}, this implies triviality
(that is vanishing of renormalized vertex functions with more than two
external lines) for $D>3$ and very likely also for $D=3$. In the
latter case triviality is manifest in the one--loop approximation and
in large$-N$ limit due to the Landau pole.  For this reason we shall
keep $\Lambda$ finite and regard the $\phi^4$ model as an effective
low--energy theory (here low--energy means practically all energies
below Planck's scale, due to the large value of the Landau pole for
renormalized coupling constants of order one or less).

We shall work in the wavefunction representation where
$\braket{\varphi}\Psi=\Psi(\varphi)$ and
\begin{equation*}
	(\phi_0\Psi)(\varphi)=\varphi_0 \Psi(\varphi)
	  \;,\quad 
	(\pi_0\Psi)(\varphi)= -i\pdif{}{\varphi_0}\Psi(\varphi)
\end{equation*}
while for $k>0$ (in lexicographic sense)
\begin{equation*}
	(\phi_{\pm k}\Psi)(\varphi)=\dfrac1{\sqrt2}\left(
	\varphi_k\pm i\,\varphi_{-k}\right)\Psi(\varphi)
	  \;,\quad 
	(\pi_{\pm k}\Psi)(\varphi)= \dfrac1{\sqrt2} \left(-i
	\pdif{}{\varphi_k}\pm \pdif{}{\varphi_{-k}}\right)\Psi(\varphi)
\end{equation*}
Notice that by construction the variables $\varphi_k$ are all real.
Of course, when either one of the cutoffs are removed, the wave function
$\Psi(\varphi)$ acquires infinitely many arguments and becomes what is
usually called a {\em wavefunctional}.

In practice, the problem of studying the dynamics of the $\phi^4$
field out of equilibrium consists now in trying to solve the
time-dependent Schroedinger equation given an initial wavefunction
$\Psi(\varphi,t=0)$ that describes a state of the field far away from
the vacuum. By this we mean a non--stationary state that, in the
infinite volume limit $L\to\infty$, would lay outside the particle
Fock space constructed upon the vacuum. This approach could be
generalized in a straightforward way to mixtures described by density
matrices, as done, for instance, in \cite{hkmp,bdv,devega3}. Here we shall
restrict to pure states, for sake of simplicity and because all
relevant aspects of the problem are already present in this case.

It is by now well known \cite{devega2} that perturbation theory is not
suitable for the purpose stated above. Due to parametric resonances
and/or spinodal instabilities there are modes of the field that grow
exponentially in time until they produce non--perturbative effects for
any coupling constant, no matter how small. On the other hand, only
few, by now standard, approximate non--perturbative schemes are
available for the $\phi^4$ theory, and to these we have to resort
after all.  We shall consider here the time-dependent Hartree--Fock
(tdHF) approach (an improved version with respect to what is
presented, for instance, in \cite{tdHF}) and the large $N$ expansion to
leading order. In fact these two methods are very closely related, as
shown for instance in \cite{inomogeneo}, where several techniques to
derive reasonable dynamical evolution equations for non--equilibrium
$\phi^4$ are compared.  However, before passing to approximations, we
would like to stress that the following rigorous result can be
immediately established in this model with both UV and IR cutoffs.

\subsection{A rigorous result: the effective potential is convex}
\label{nogo}

This is a well known fact in statistical mechanics, being directly
related to stability requirements. It would therefore hold also for
the field theory in the Euclidean functional formulation. In our
quantum--mechanical context we may proceed as follow.  Suppose the
field $\phi$ is coupled to a uniform external source $J$.  Then the
ground state energy $E_0(J)$ is a concave function of $J$, as can be
inferred from the negativity of the second order term in $\Delta J$ of
perturbation around any chosen value of $J$. Moreover, $E_0(J)$ is
analytic in a finite neighborhood of $J=0$, since $J\phi$ is a
perturbation ``small'' compared to the quadratic and quartic terms of
the Hamiltonian. As a consequence, this effective potential
$V_{\text{eff}}(\bar\phi)=E_0(J)-J\bar\phi$,
$\bar\phi=E_0'(J)=\vev{\phi}_0$, that is the Legendre transform of
$E_0(J)$, is a convex analytic function in a finite neighborhood of
$\bar\phi=0$.  In the infrared limit $L\to\infty$, $E_0(J)$ might
develop a singularity in $J=0$ and $V_{\text{eff}}(\bar\phi)$ might
flatten around $\bar\phi=0$. Of course this possibility would apply in
case of spontaneous symmetry breaking, that is for a double--well
classical potential. This is a subtle and important point that will
play a crucial role later on, even if the effective potential is
relevant for the static properties of the model rather than the
dynamical evolution out of equilibrium that interests us here. In fact
such evolution is governed by the CTP effective action
\cite{schwinger,zshy} and one might expect that, although non--local
in time, it asymptotically reduces to a multiple of the effective
potential for trajectories of $\bar\phi(t)$ with a fixed point at
infinite time. In such case there should exist a one--to--one
correspondence between fixed points and minima of the effective
potential. This is one of the topics addressed in this paper.

\section{Large $N$ expansion at leading order}\label{ln}
\subsection{Definitions}
In this section we consider a standard non--perturbative approach to
the $\phi^4$ model which is applicable also out of equilibrium, namely
the large $N$ method as presented in \cite{chkmp}. However we shall follow
a different derivation which makes the gaussian nature of the 
$N\to\infty$ limit more explicit.

In the large $N$ method one generalizes the $\phi^4$ model by
promoting the single real scalar field $\phi$ to a $N-$component
vector $\bds\phi$ of scalar fields, in such a way to ensure $O(N)$
symmetry. This corresponds to the hamiltonian
\begin{equation*}
\begin{split}
  H &=\frac12\int d^Dx \left[\bds\pi^2 + (\partial\bds\phi)^2 +
  \mbare\,\bds\phi^2 + \lbare( \bds\phi^2)^2 \right] = \\
    &=\frac12\sum_k\left[\bds\pi_k \!\cdot \bds\pi_{-k} +(k^2+
  \mbare)\,\bds\phi_k \!\cdot \bds\phi_{-k}\right]
	+\dfrac\lbare{4L^D} \sum_{k_1,k_2,k_3,k_4} (\bds\phi_{k_1}\!\!\cdot
  \bds\phi_{k_2})(\bds\phi_{k_3} \!\!\cdot \bds\phi_{k_4}) 
	\,\delta^{(D)}_{k_1+k_2+k_3+k_4,0}
\end{split}
\end{equation*}
where the space integration and the sum over wavevectors are limited
by the infrared and ultraviolet cutoff, respectively, according to the
general framework presented in section \ref{cft}.

It is known that this theory is well behaved for large $N$, provided
that the quartic coupling constant $\lambda$ is rescaled with
$1/N$. For example, it is possible to define a perturbation theory,
based on the small expansion parameter $1/N$, in the framework of
which one can compute any quantity at any chosen order in $1/N$. From
the diagrammatic point of view, this procedure corresponds to a
resummation of the usual perturbative series that automatically
collects all the graphs of a given order in $1/N$ together
\cite{largen_exp}. Moreover, it has been established since the early 80's
that the leading order approximation (that is the strict limit
$N\to\infty$) is actually a classical limit \cite{yaffe}, in the sense
that there exists a classical system (i.e., a classical phase space, a
Poisson bracket and a classical Hamiltonian) whose dynamics controls
the evolution of all fundamental quantum observables, such as field
correlation functions, in the $N\to\infty$ limit. For instance, from
the absolute minimum of the classical Hamiltonian one reads the energy
of the ground state, while the spectrum is given by the frequencies of
small oscillations about this minimum, etc. etc.. We are here
interested in finding an efficient and rapid way to compute the
quantum evolution equations for some observables in the $N \to 
\infty$ limit, and we will see that this task is easily accomplished
just by deriving the canonical Hamilton equations from the large $N$
classical Hamiltonian.

Following Yaffe \cite{yaffe}, we write the quantum mechanical
hamiltonian as 
\begin{equation}\label{hAC}
	H = N h (A\,,C) 
\end{equation}
in terms of the square matrices $A$, $C$ with operator entries
({\bf $\varpi_k$} is the canonical momentum conjugated to the real mode
{\bf $\varphi_k$})
\begin{equation*}
	A_{kk'} = \frac1{N} \bds\varphi_k \!\cdot \bds\varphi_{k'}
   \;,\quad C _{kk'} = \frac1{N} \bds\varpi_k\!\cdot \bds\varpi_{k'}
\end{equation*}
These are example of ``classical'' operators, whose two-point correlation
functions factorize in the $N\to\infty$ limit. This can be shown by
considering the coherent states 
\begin{equation}
	\Psi_{z,q,p}( \bds\varphi) = C(z) \exp\left[ 
	i\sum_k \bds p_k\cdot \bds\varphi_k - \frac1{2N} \sum_{kk'} z_{kk'} 
	(\bds\varphi_k-\bds q_k) \cdot(\bds\varphi_{k'}-\bds q_{k'}) \right]
\label{coh_states}
\end{equation}
where the complex symmetric matrix $z$ has a positive definite real
part while $\bds p_k$ and $\bds q_k$ are real and coincide,
respectively, with the coherent state expectation values of $\bds
\varpi_k$ and $\bds \varphi_k$. As these parameters take  all their
possible values, the coherent states form an overcomplete set in the
cutoff Hilbert space of the model. The crucial property which ensures
factorization is that they become all orthogonal in the $N\to\infty$
limit. Moreover one can show \cite{yaffe} that the 
coherent states parameters form a classical phase space with Poisson
brackets 
\begin{equation*}
	\left\{q_k^i\,,p_{k'}^j\right\}_{\rm P.B.} = \delta_{kk'}
	\delta^{ij} \;,\quad
	\left\{w_{kk'}\,,v_{qq'}\right\}_{\rm P.B.} = 
	\delta _{kq} \delta _{k'q'} + \delta _{kq'} \delta _{k'q}
\end{equation*}
where $w$ and $v$ reparametrize $z$ as $z=\frac12
w^{-1}+i\,v$. It is understood that the dimensionality of the vectors
{\bf $q_k$} and {\bf $p_k$} is arbitrary but finite [that is, only a finite
number, say $n$, of pairs $(\varphi_k^i\,,\varpi_k^i)$ may take a nonvanishing
expectation value as $N\to\infty$].

Once applied to the classical operators $A_{kk'}$ and
$C_{kk'}$ the large $N$ factorization allow to obtain the classical
hamiltonian by simply replacing $A$ and $C$ in eq. (\ref{hAC}) by the
coherent expectation values
\begin{equation*}
	\vev{A_{kk'}} = \bds q_k\cdot \bds q_{k'} + w_{kk'} \;,\quad
	\vev{C_{kk'}} = \bds p_k\cdot \bds p_{k'} + 
	(v\,w\,v)_{kk'} + \frac14 ( w^{-1})_{kk'}
\end{equation*}
In our situation, having assumed a uniform background expectation
value for $\phi$, we have $\bds q_k=\bds p_k=0$ for all
$k\neq 0$; moreover, translation invariance implies that 
$w$ and $v$ are diagonal matrices, so that we may set
\begin{equation*}
	w_{kk'} = \s_k^2\,\delta_{kk'} \;,\quad 
	v_{kk'}=\frac{s_k}{\s_k}\,\delta_{kk'}
\end{equation*}
in term of the canonical couples $(\s_k,s_k)$ which satisfy 
$\{\s_k\,,s_{k'}\}_{\rm P.B.}=\delta_{kk'}$. Notice that the  $\s_k$ are
just the widths (rescaled by $N^{-1/2}$) of the $O(N)$ symmetric
and translation invariant gaussian coherent states.

Thus we find the classical hamiltonian
\begin{equation*}
	h_{\rm cl} =\frac12(\bds p_0^2 +\mbare\, \bds q^2_0) +
	\frac12\sum_k\left[ s_k^2 +(k^2+\mbare)\s_k^2 + \frac1{4\s^2_k}
	\right] + \frac\lbare{4L^D} \left(\bds q_0^2+\sum_k\s_k^2\right)^2
\end{equation*}
where by Hamilton's equations of motion $\bds p_0=\dot{\bds q}_0$ and
$s_k=\dot\s_k$. The corresponding conserved energy density
$\E=L^{-D}h_{\rm cl}$ may be written
\begin{equation}\label{ElargeN}
\begin{split}
	\E &= \T +\V \;,\quad \T = \frac12\dot{\bar{\bds\phi}}^2 +
	\frac1{2L^D}\sum_k \dot\s_k^2  \\ \V &= \frac1{2L^D}\sum_k\left(
	 k^2\,\s_k^2 + \frac1{4\s^2_k}\right) +V(\bar{\bds\phi}^2
	+\Sigma) \;,\quad \Sigma=\frac1{L^D}\sum_k\s_k^2
\end{split}
\end{equation}
where $\bar{\bds\phi}=L^{-D/2}\bds q_0$ and $V$ is
the $O(N)-$invariant quartic potential regarded as a function of
$\bds\phi^2$, that is $V(z)=\frac12\mbare z + \frac14\lbare z^2$. It
is worth noticing that eq. (\ref{ElargeN}) would apply as is to
generic $V(z)$.

\subsection{Static properties}
Let us consider first the static aspects embodied in the effective
potential $V_{\text{eff}}(\bar{\bds\phi})$, that is the minimum of the
potential energy $\V$ at fixed $\bar{\bds\phi}$.  We first define in a
precise way the unbroken symmetry phase, in this large $N$ context, as
the case when $V_{\text{eff}}(\bar{\bds\phi})$ has a unique minimum at
$\bar{\bds\phi}=0$ in the limit of infinite volume. Minimizing $\V$
w.r.t. $\s_k$ yields
\begin{equation}\label{massive}
\begin{split}
	\s^2_k=\frac1{2\sqrt{k^2+M^2}} \;,\quad
	M^2 &= \mbare + 2\,V'(\bar{\bds\phi}^2+\Sigma) \\
	&= \mbare + \lbare\bar{\bds\phi}^2+ \frac\lbare{L^D}
	\sum_k\frac1{2\sqrt{k^2+M^2}}
\end{split}
\end{equation}
that is the widths characteristic of a free theory with
self--consistent mass $M$ fixed by the gap equation. By the assumption
of unbroken symmetry, when $\bar{\bds\phi}=0$ and at infinite volume
$M$ coincides with the equilibrium mass $m$ of the theory, that may be
regarded as independent scale parameter. Since in the limit
$L\to\infty$ sums are replaced by integrals
\begin{equation*}
	\Sigma \to \int_{k^2\le\Lambda^2} \dfrac{d^Dk}{(2\pi)^D} \s_k^2
\end{equation*}
we obtain the standard bare mass parameterization
\begin{equation}\label{Nm2ren}
	\mbare= m^2 - \lbare I_D(m^2,\Lambda) \;,\quad 
	I_D(z,\Lambda) \equiv \int_{k^2\le\Lambda^2} 
	\dfrac{d^Dk}{(2\pi)^D} \dfrac1{2\sqrt{k^2+z}}
\end{equation}
and the renormalized gap equation
\begin{equation}\label{rengapN}
	M^2 = m^2+ \l\,\bar\phi^2+\l \left[ I_D(M^2,\Lambda)-
	I_D(m^2,\Lambda) \right]_{\text{finite}} 
\end{equation}
which implies, when $D=3$,
\begin{equation}\label{lrenN}
	\lbare =\l \left(1-\frac{\l}{8\pi^2}
	    \log\frac{2\Lambda}{m\sqrt{e}} \right)^{\!\!-1} 
\end{equation}
with a suitable choice of the finite part. No coupling constant
renormalization occurs instead when $D=1$. The renormalized gap
equation (\ref{rengapN}) may also be written quite concisely
\begin{equation}\label{nice}
	\frac{M^2}{\hat\l(M)} = \frac{m^2}{\hat\l(m)} + \,\bar{\bds\phi}^2
\end{equation}
in terms of the one--loop running couplings constant
\begin{equation*}
	\hat\l(\mu) = \l \left[ 1 - \frac{\l}{8\pi^2} 
	\log\frac{\mu}m \right]^{-1} \;,\quad \hat\l(m)=\l
	\;,\quad \hat\l(2\Lambda\,e^{-1/2}) =\lbare
\end{equation*}
It is the Landau pole in $\hat\l(2\Lambda\,e^{-1/2})$ that actually
forbids the limit $\Lambda\to\infty$. Hence we must keep the cutoff
finite and smaller than $\Lambda_{\text{pole}}$, so that the theory
does retain a slight inverse--power dependence on it. At any rate,
there exists a very wide window where this dependence is indeed very
weak for couplings of order one or less, since
$\Lambda_{\text{pole}}=\frac12 m\exp(1/2+8\pi^2/\l)\gg m$.  Moreover,
we see from eq. (\ref{nice}) that for $\sqrt\l|\bar\phi|$ much smaller
than the Landau pole there are two solutions for $M$, one
``physical'', always larger than $m$ and of the same order of
$m+\sqrt\l|\bar\phi|$, and one ``unphysical'', close to the Landau
pole.

One can now easily verify that the effective potential has indeed a
unique minimum in $\bar{\bds\phi}=0$, as required. In fact, if we
assign arbitrary $\bar\phi-$dependent values to the widths $\s_k$,
(minus) the effective force reads
\begin{equation}\label{Nefforce}
	\der{}{\bar\phi^i}\V(\bar{\bds\phi},\{\s_k(\bar{\bds\phi})\})
	= M^2\,\bar\phi^i + \sum_k \pdif\V{\s_k}
	\der{\s_k}{\bar\phi^i}
\end{equation} 
and reduces to $M^2\,\bar\phi^i$ when the widths are extremal as
in eq. (\ref{massive}); but $M^2$ is positive for unbroken symmetry
and so $\bar{\bds\phi}=0$ is the unique minimum. 

We define the symmetry as broken whenever the infinite volume
$V_{\text{eff}}$ has more than one minimum.  Of course, as long as $L$
is finite, $V_{\text{eff}}$ has a unique minimum in
$\bar{\bds\phi}=0$, because of the unicity of the ground state in
Quantum Mechanics, as already discussed in section \ref{nogo}. Let us
therefore proceed more formally and take the limit $L\to\infty$
directly on the potential energy $\V$. It reads
\begin{equation*}
	\V = \frac12 \int_{k^2\le\Lambda^2} \dfrac{d^Dk}{(2\pi)^D}
	 \left(k^2\,\s_k^2 + \frac1{4\s^2_k}\right) 
	+V(\bar{\bds\phi}^2 +\Sigma) \;,\quad  
	\Sigma =\int_{k^2\le\Lambda^2} \dfrac{d^Dk}{(2\pi)^D}\,\s_k^2
\end{equation*}
where we write for convenience the tree--level potential $V$ in the
positive definite form $V(z)=\frac14\lbare(z+\mbare/\lbare)^2$.
$\V$ is now the sum of two positive definite terms.
Suppose there exists a configuration such that 
$V(\bar{\bds\phi}^2 +\Sigma)=0$ and the first term in $\V$ is at
its minimum. Then this is certainly the absolute minimum of $\V$.
This configuration indeed exists at infinite volume when $D=3$:
\begin{equation}\label{minimum}
	\s^2_k=\frac1{2|k|} \;,\quad  \bar{\bds\phi}^2 =v^2
	\;,\quad \mbare = -\lbare\left[ v^2+ I_3(0,\Lambda) \right]
\end{equation}
where the nonnegative $v$ should be regarded as an independent
parameter fixing the scale of the symmetry breaking. It replaces the
mass parameter $m$ of the unbroken symmetry case: now the theory is
massless in accordance with Goldstone theorem.  On the contrary, if
$D=1$ this configuration is not allowed due to the infrared
divergences caused by the massless nature of the width spectrum. This
is just the standard manifestation of Mermin--Wagner--Coleman theorem
that forbids continuous symmetry breaking in a two--dimensional
space--time \cite{mwc}.

At finite volumes we cannot minimize the first term in $\V$ since this
requires $\s_0$ to diverge, making it impossible to keep
$V(\bar{\bds\phi}^2 +\Sigma)=0$. In fact we know that the
unicity of the ground state with finitely many degrees of freedom
implies the minimization equations (\ref{massive}) to hold always
true with a $M^2$ strictly positive.  Therefore, broken symmetry
should manifest itself as the situation in which the equilibrium value
of $M^2$ is a positive definite function of $L$ which vanishes in the
$L\to\infty$ limit.

We can confirm this qualitative conclusion as follows. We assume that
the bare mass has the form given in eq. (\ref{minimum}) and that 
$\bar{\bds\phi}^2 =v^2$ too. Minimizing the potential energy
leads always to the massive spectrum, eq. (\ref{massive}), with the
gap equation
\begin{equation}\label{gapagain}
	\frac{M^2}\lbare = \frac1{2L^3M} + \frac1{2L^3}
	\sum_{k\neq0}\frac1{\sqrt{k^2+M^2}} - \frac{\Lambda^2}{8\pi^2}
\end{equation}
If $M^2>0$ does not vanish too fast for large volumes, or stays even
finite, then the sum on the modes has a behavior similar to the
corresponding infinite volume integral: there is a quadratic
divergence that cancels the infinite volume contribution, and a
logarithmic one that renormalizes the bare coupling. The direct
computation of the integral would produce a term containing the
$M^2\log(\Lambda/M)$. This can be split into
$M^2[\log(\Lambda/v)-\log(M/v)]$ by using $v$ as mass scale. The first
term renormalizes the coupling correctly, while the second one
vanishes if $M^2$ vanishes in the infinite volume limit.

When $L\to\infty$, the asymptotic solution of (\ref{gapagain}) reads
\begin{equation*}
	M = \left( \frac{\l}{2}\right) ^{1/3} L ^{-1} + \rm{h.o.t.}
\end{equation*}
that indeed vanishes in the limit. Note also that the exponent is
consistent with the assumption made above that $M$ vanishes slowly
enough to approximate the sum over $k\neq 0$ with an integral with the
same $M$.

Let us now consider a state whose field expectation value
$\bar{\bds\phi}^2$ is different from $v^2$. If $\bar{\bds\phi}^2 > v$,
the minimization equations (\ref{massive}) leads to a positive squared
mass spectrum for the fluctuations, with $M^2$ given
self--consistently by the gap equation. On the contrary, as soon as
$\bar{\bds\phi}^2 < v^2$, one immediately see that a positive $M^2$
cannot solve the gap equation
\begin{equation*}
	M^2 = \lbare \left( \bar{\bds\phi}^2 -v^2 + \frac{\s_0^2}{L^3} +
	\frac1{2L^3} \sum_{k\neq0}\frac1{\sqrt{k^2+M^2}} - 
	\frac{\Lambda^2}{8 \pi ^2} \right)
\end{equation*}
if we insist on the requirement that $\s_0$ not be macroscopic. In fact,
the r.h.s. of the previous equation is negative, no matter which
positive value for the effective mass we choose, at least for $L$
large enough. But nothing prevent us to consider a static
configuration for which the amplitude of the zero mode is
macroscopically large (i.e. it rescales with the volume $L^3$).
Actually, if we choose
\begin{equation*}
	\frac{\s _0 ^2}{L^3} =  v ^2 - \bar{\bds\phi}^2 + \frac1{2L^3M}
\end{equation*}
we obtain the same equation as we did before and the same value for
the potential, that is the minimum, in the limit $L \to \infty$. Note
that at this level the effective mass $M$ needs not to have the same
behavior in the $L \to \infty$ limit, but it is free of rescaling
with a different power of $L$. We can be even more precise: we isolate
the part of the potential that refers to the zero mode width $\s_0$
($\Sigma'$ does not contain the $\s_0$ contribution)
\begin{equation*}
\frac12 \left[ \mbare + \lbare \left( \bar{\bds\phi}^2 +\Sigma' \right)
\right] \frac{\s _0 ^2}{L ^3} + \frac{\lbare}4 \frac{\s _0 ^4}{L ^6} +
\frac{1}{8 L ^3 \sigma _0 ^2}
\end{equation*}
and we minimize it, keeping $\bar{\bds\phi}^2$ fixed. The minimum is
attained at $t=\s_0^2/L^3$ solution of the cubic equation
\begin{equation*}
\lbare t^3 + \a \lbare t^2 - \tfrac14 L^{-6} = 0
\end{equation*}
where $\a = \bar{\bds\phi}^2 - v ^2 + \Sigma ' - I _3 \left( 0 ,
\Lambda \right)$. Note that $\lbare \alpha$ depends on $L$ and it has
a finite limit in infinite volume: $\l (\bar{\bds\phi}^2 - v ^2)$. The
solution of the cubic equation is
\begin{equation*}
	\lbare t = \lbare ( v ^2 - \bar{\bds\phi}^2 ) + 
	\tfrac14 [L^3(v^2 -\bar{\bds\phi}^2)]^{-2} + \rm{h.o.t.}
\end{equation*}
from which the effective mass can be identified as proportional
to $L^{-3}$. The stability equations for all the other modes can now
be solved by a massive spectrum, in a much similar way as before. 

Since $\s_0$ is now macroscopically large, the infinite volume limit
of the $\s_k$ distribution (that gives a measure of the {\em transverse}
fluctuations in the $O(N)$ model) develop a $\delta-$like singularity, 
signalling  a Bose condensation of the Goldstone bosons: 
\begin{equation}\label{BE}
	\s_k^2 = ( v ^2 - \bar{\bds\phi}^2 )\,\delta^{(D)}(k)+
	\frac1{2k}
\end{equation}
At the same time it is evident that the minimal potential energy is
the same as when $\bar{\bds\phi}^2=v^2$, that is  the effective
potential flattens, in accord with the Maxwell construction.

Eq. (\ref{BE}) corresponds in configuration space to the $2-$point
correlation function
\begin{equation}\label{2pN}
	\lim_{N \to \infty} \frac{\vev{\bds{\phi}(x)
	\cdot\bds{\phi}(y)}}{N} = \bar{\bds{\phi}}^2+ \int\frac{d^Dk}
	{(2\pi)^D}\,\s_k^2\, e^{ik\cdot(x-y)}= C(\bar{\bds{\phi}}^2)
	+\Delta_D(x-y)
\end{equation}
where $\Delta_D(x-y)$ is the massless free--field equal--time
correlator, while  
\begin{equation}\label{broken}
	C(\bar{\bds{\phi}}^2)= v^2\,\Theta(v^2-\bar{\bds{\phi}}^2) +
	\bar{\bds{\phi}}^2\, \Theta(\bar{\bds{\phi}}^2-v^2) =
	\text{max}(v^2,\bar{\bds{\phi}}^2)
\end{equation}
This expression can be extended to unbroken symmetry by
setting in that case $C(\bar{\bds{\phi}}^2)=\bar{\bds{\phi}}^2$.

Quite evidently, when eq. (\ref{broken}) holds, symmetry breaking can
be inferred from the limit $|x-y|\to\infty$, if clustering is assumed
\cite{zj,allthat}, since $\Delta_D(x-y)$ vanishes for large
separations. Of course this contradicts the infinite volume limit of
the finite--volume definition, $\bar{\bds\phi}=\lim_{N\to\infty}
N^{-1/2}\vev{\bds\phi(x)}$, except at the extremal points
$\bar{\bds{\phi}}^2=v^2$.

In fact the $L\to\infty$ limit of the finite volume states with
$\bar{\bds\phi}^2<v^2$ do violate clustering, because they are linear
superpositions of vectors belonging to superselected sectors and
therefore they are indistinguishable from statistical mixtures. We
shall return in more detail on this aspects in section \ref{tdhf},
where a generalized HF approximation is considered for $N=1$. For the
moment we can give the following intuitive picture for large
$N$. Consider any one of the superselected sectors based on a physical
vacuum with $\bar{\bds\phi}^2=v^2$. By condensing a macroscopic number
of transverse Goldstone bosons at zero--momentum, one can build
coherent states with rotated $\bar{\bds\phi}$. By incoherently
averaging over such rotated states one obtains new states with
field expectation values shorter than $v$ by any prefixed amount.
In the large $N$ approximation this averaging is necessarily uniform
and is forced upon us by the residual $O(N-1)$ symmetry. 

\subsection{Out--of--equilibrium dynamics}\label{ooedN}

We now turn to the dynamics out of equilibrium in this large $N$
context. It is governed by the equations of motion derived from the
total energy density $\E$ in eq. (\ref{ElargeN}), that is
\begin{equation}\label{sk2}
	\der{^2\bar{\bds\phi}}{t^2} = -M^2\,\bar{\bds\phi}
	\;,\quad \der{^2 \s_k}{t^2}= -(k^2+M^2)\,\s_k + \frac1{4 \s_k^3}
\end{equation}
where the generally time--dependent effective squared mass $M^2$ is
given by
\begin{equation}\label{Nunbgap}
	M^2 = m^2 + \lbare
	\left[\bar{\bds\phi}^2+\Sigma-I_D(m^2,\Lambda)\right]
\end{equation}
in case of unbroken symmetry and
\begin{equation}\label{Nbgap}
	M^2 = \lbare \left[\bar{\bds\phi}^2- v^2 +\Sigma-I_3(0,\Lambda)\right]
\end{equation}
for broken symmetry in $D=3$.

At time zero, the specific choice of initial conditions for $\s_k$ that
give the smallest energy contribution, that is
\begin{equation}\label{inis}
	\dot \s_k=0 \;,\quad \s_k^2=\frac1{2\sqrt{k^2+M^2}}
\end{equation}
turns eq. (\ref{Nunbgap}) into the usual gap equation
(\ref{massive}). For any value of $\bar\phi$ this equation has one
solution smoothly connected to the value $M=m$ at $\bar\phi=0$.  Of
course other initial conditions are possible. The only requirement is
that the corresponding energy must differ from that of the ground
state by an ultraviolet finite amount, as it occurs for the choice
(\ref{inis}). In fact this is guaranteed by the gap equation
itself, as evident from eq. (\ref{Nefforce}): when the widths $\s_k$ 
are extremal the effective force is finite, and therefore so are all
potential energy differences.

This simple argument needs a refinement in two respects. 

Firstly, in case of symmetry breaking the formal energy minimization
w.r.t. $\s_k$ leads always to eqs. (\ref{inis}), but these are
acceptable initial conditions only if the gap equation that follows
from eq. (\ref{Nbgap}) in the $L\to\infty$ limit, namely
\begin{equation}\label{Nbgap2}
	M^2 = \lbare \left[\bar{\bds\phi}^2-v^2 +
	I_D(M^2,\Lambda)-I_D(0,\Lambda) \right] 
\end{equation}
admits a nonnegative, physical solution for $M^2$. 

Secondly, ultraviolet finiteness only requires that the sum over $k$
in eq. (\ref{Nefforce}) be finite and this follows if eq. (\ref{inis})
holds at least for $k$ large enough, solving the issue raised in the
first point: negative $M^2$ are allowed by imposing a new form of gap
equation
\begin{equation}\label{newgap}
	M^2 = \lbare \left[\bar{\bds\phi}^2-v^2 + \frac1{L^D} 
	\sum_{k^2<|M^2|}\s_k^2 + \frac1{L^D}\sum_{k^2>|M^2|}
	\frac1{2\sqrt{k^2-|M^2|}} -I_D(0,\Lambda) \right] 
\end{equation}
where all $\s_k$ with $k^2<|M^2|$ are kept free (but all by hypothesis
microscopic) initial conditions. Of course there is no energy
minimization in this case. To determine when this new form is
required, we observe that, neglecting the inverse--power corrections
in the UV cutoff we may write eq. (\ref{Nbgap2}) in the following
form 
\begin{equation}\label{NiceN}
	\frac{M^2}{\hat\l(M)} = \bar{\bds\phi}^2 -v^2
\end{equation}
There exists a positive solution $M^2$ smoothly connected to the
ground state, $\bar{\bds\phi}^2=v^2$ and $M^2=0$, only provided
$\bar{\bds\phi}^2\ge v^2$. So, in the large $N$ limit, as soon as we start
with $\bar{\bds\phi}^2\le v^2$, we cannot satisfy the gap equation with a
positive value of $M^2$. The situation will be quite different in the
case of $N=1$ in the HF approximation (cfr. section \ref{ooed}).

Once a definite choice of initial conditions is made, the system of
differential equations (\ref{sk2}), (\ref{Nunbgap}) or (\ref{Nbgap})
can be solved numerically with standard integration algorithms.  This
has been already done by several authors \cite{bdvhs,relax,devega2}, working
directly in infinite volume, with the following general results.  In
the case of unbroken symmetry it has been established that the $\s_k$
corresponding to wavevectors $k$ in the so--called forbidden bands
with parametric resonances grow exponentially in time until their
growth is shut off by the back--reaction. For broken symmetry it is
the region in $k-$space with the spinodal instabilities caused by an
initially negative $M^2$, whose widths grow exponentially before the
back--reaction shutoff.  After the shutoff time the effective mass
tends to a positive constant for unbroken symmetry and to zero for
broken symmetry (in D=3), so that the only width with a chance to keep
growing indefinitely is $\s_0$ for broken symmetry.

Of course, in all these approaches the integration over modes in the
back--reaction $\Sigma$ cannot be done exactly and is always replaced
by a discrete sum of a certain type, depending on the details of the
algorithms. Hence there exists always an effective infrared cutoff,
albeit too small to be detectable in the numerical outputs.  A
possible troublesome aspect of this is the proper identification of
the zero--mode width $\s_0$. Even if a (rather arbitrary) choice of
discretization is made where a $\s_0$ appears, it is not really
possible to determine whether during the exponential growth or after
such width becomes of the order of the volume.  Our aim is just to
answer this question and therefore we perform our numerical
evolution in finite volumes of several growing sizes.
Remanding to the appendix for the details of our method, we summarize
our results in the next subsection.

\subsection{Numerical results}
\begin{figure}
\epsfig{file=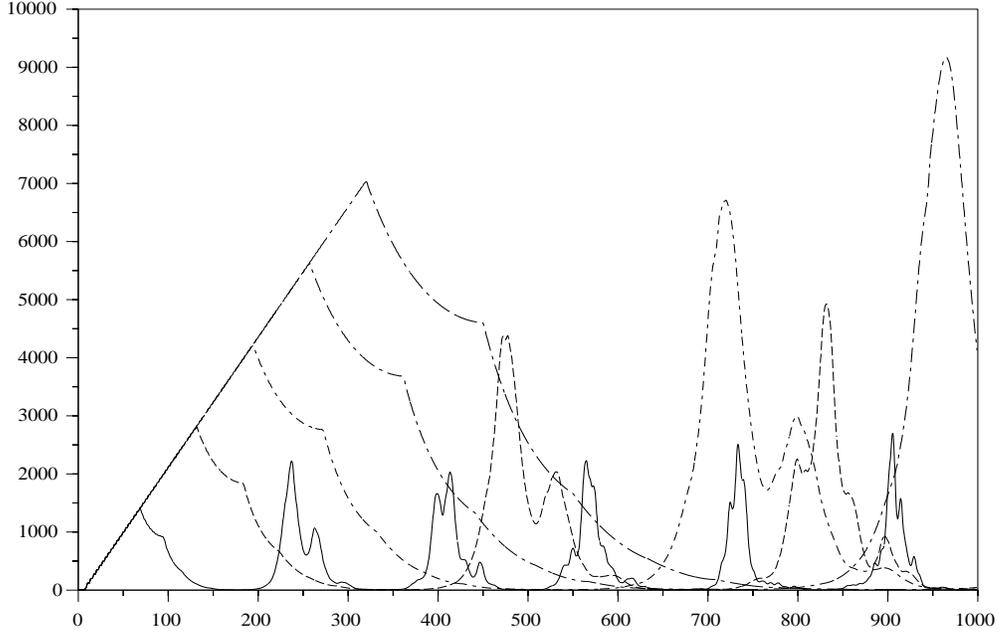,height=10cm,width=15cm}
\caption{\it Zero--mode amplitude evolution for 
different values of the size
$L=20,40,60,80,100$, for $\lambda = 0.1$ and broken symmetry,
with $\bar\phi=0$. }\label{fig:m0}
\end{figure}

\begin{figure}
\epsfig{file=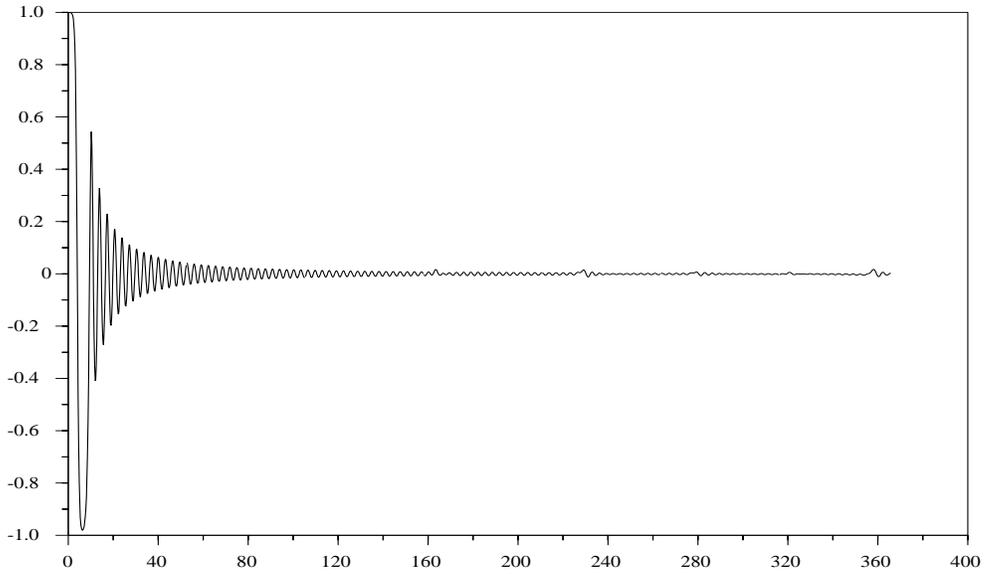,height=9cm,width=15cm}
\caption{\it Time evolution of the squared effective mass $M^2$ in broken
symmetry, for $L=100$ and $\lambda=0.1$. }\label{fig:mass2}
\end{figure}

\begin{figure}
\epsfig{file=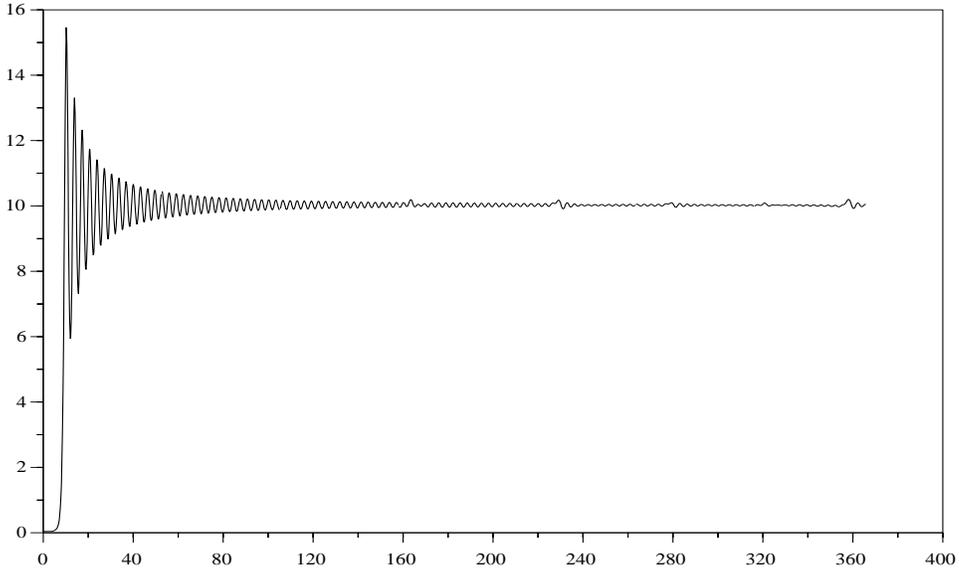,height=9cm,width=15cm}
\caption{\it The quantum back--reaction $\Sigma$, with the parameters as
in Fig. \ref{fig:mass2} }\label{fig:sigma}
\end{figure}

\begin{figure}
\epsfig{file=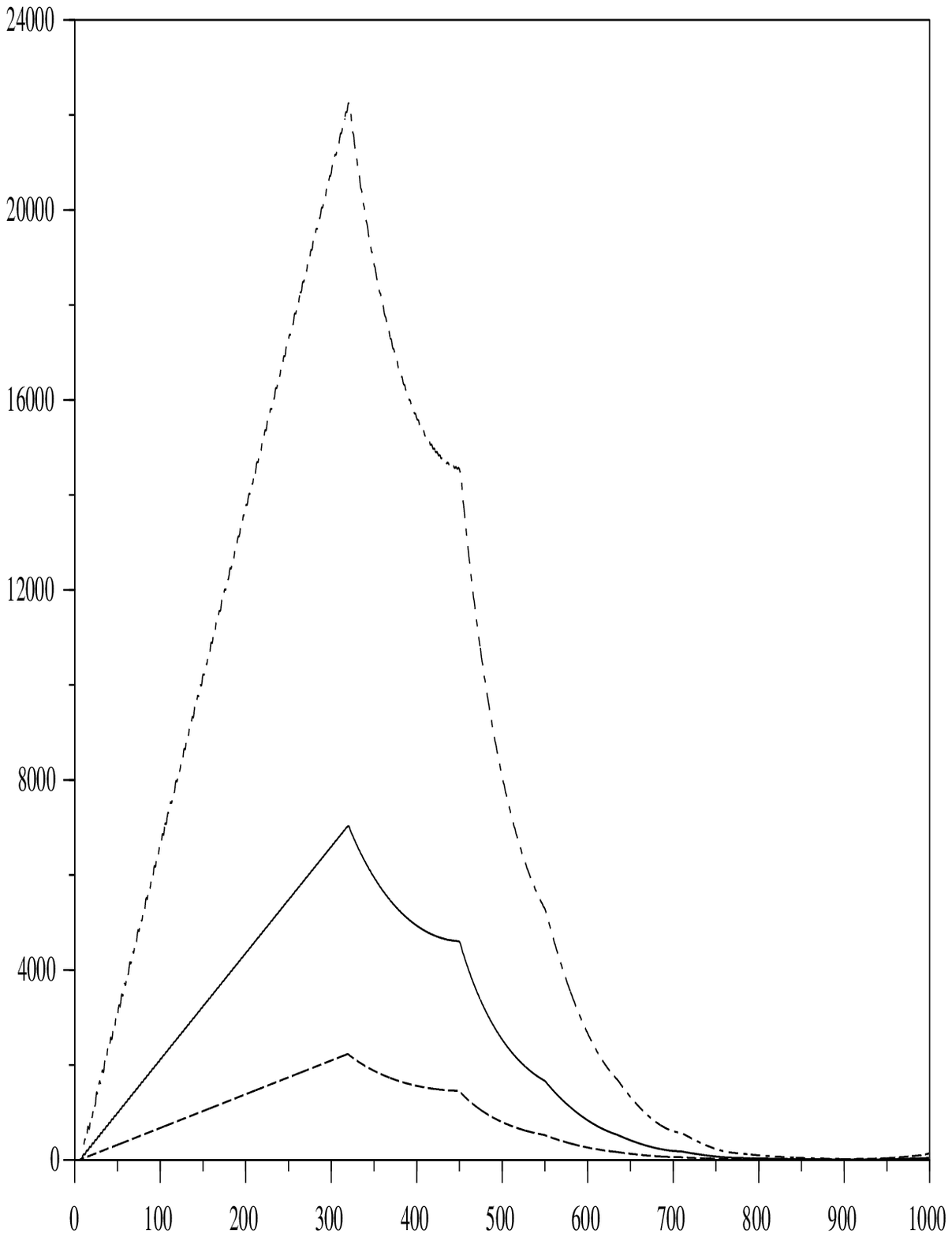,height=9cm,width=15cm}
\caption{\it Zero--mode amplitude evolution for different value of the
renormalized coupling constant $\l=0.01,0.1,1$, for $L=100$ and
broken symmetry, with $\bar\phi=0$. }\label{fig:m0_l}
\end{figure}

\begin{figure}
\epsfig{file=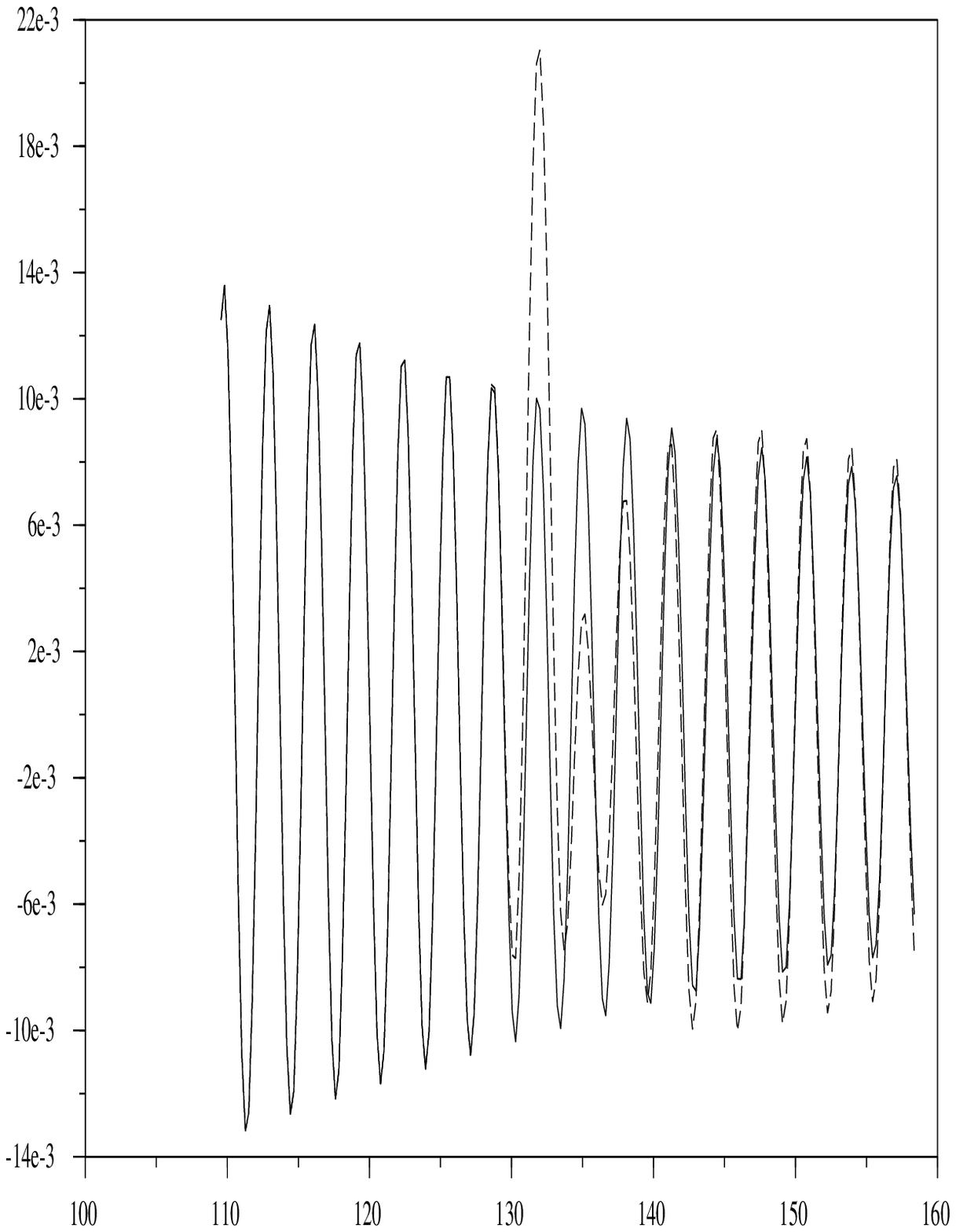,height=9cm,width=15cm}
\caption{\it Detail of $M^2$ near $t=\tau_L$ for $L=40$ (dotted
line). The case $L=80$ is plotted for comparison (solid line).}\label{fig:usc_mass}
\end{figure}

\begin{figure}
\epsfig{file=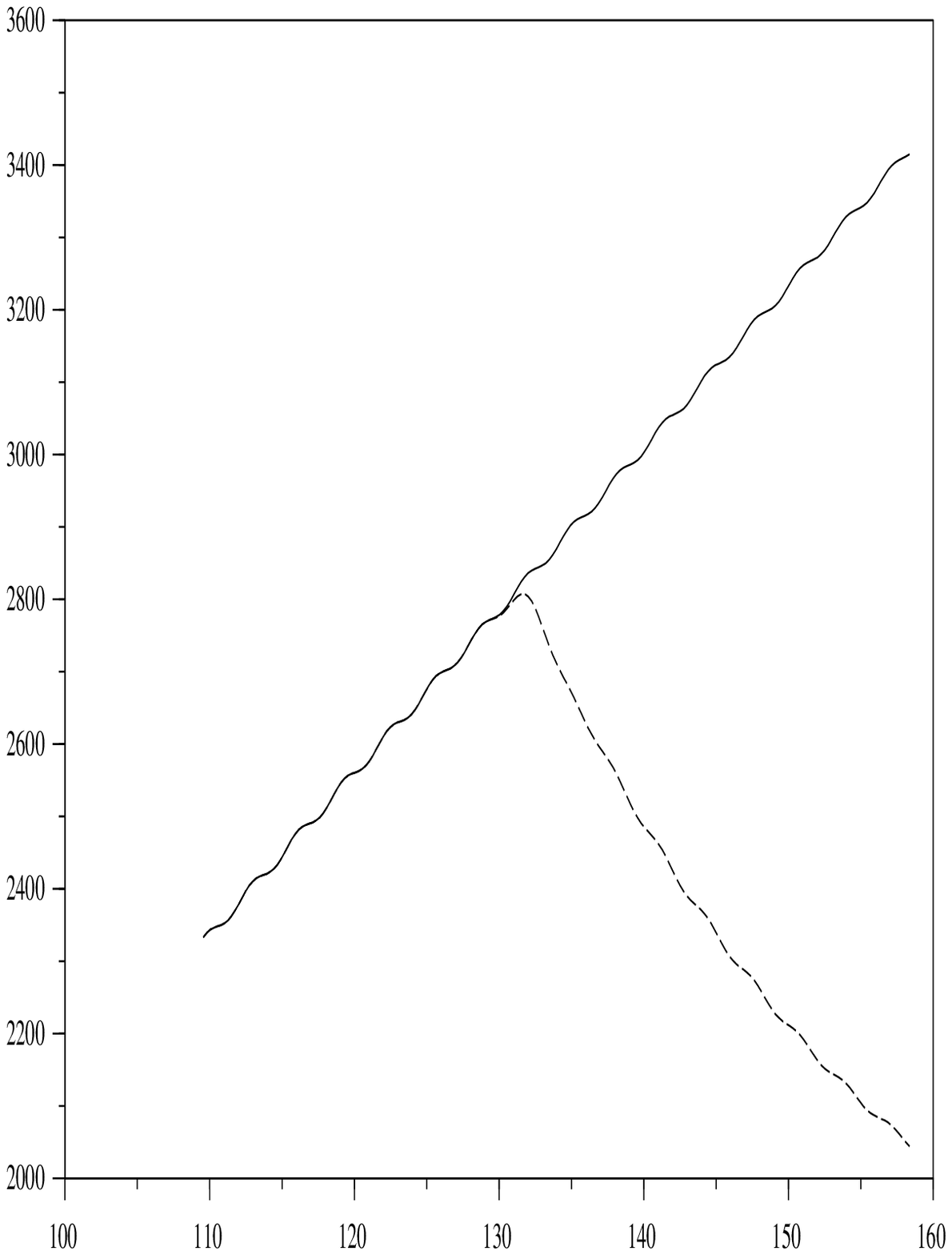,height=9cm,width=15cm}
\caption{\it Detail of $\s_0$ near $t=\tau_L$ for $L=40$ (dotted
line). The case $L=80$ is plotted for comparison (solid line).}\label{fig:usc_zm}
\end{figure}

After a careful study in $D=3$ of the scaling behavior of the
dynamics with respect to different values of $L$, the linear size of
the system, we reached the following conclusion: there exist a
$L-$dependent time, that we denote by $\tau_L$, that splits
the evolution in two parts; for $t \leq \tau_L$, the
behavior of the system does not differ appreciably from its
counterpart at infinite volume, while finite volume effects abruptly
alter the evolution as soon as $t$ exceeds $\tau_L$;
in particular
\begin{itemize}
\item $\tau_L$ is proportional to the linear size of the
box $L$ and so it rescales as the cubic root of the volume.
\item $\tau_L$ does not depend on the value of the quartic
coupling constant $\lambda$, at least in a first approximation.
\end{itemize}
The figures show the behavior of the width of the zero mode $\s_0$
(see Fig. \ref{fig:m0}), of the squared effective mass $M^2$ (see Fig.
\ref{fig:mass2} ) and of the back--reaction $\Sigma$ (see
Fig. \ref{fig:sigma}), in the more interesting case of broken
symmetry. The initial conditions are chosen in several different ways
(see the appendix for details), but correspond to a negative $M^2$ at
early times with the initial widths all microscopic, that is at most
of order $L^{1/2}$. This is particularly relevant for the zero--mode
width $\s_0$, which is instead macroscopic in the lowest energy state
when $\bar{\bds\phi}^2<v^2$, as discussed above. As for the
background, the figures are relative to the simplest case
$\bar{\bds\phi}= 0 =\dot{\bar{\bds\phi}}$, but we have considered also
initial conditions with $\bar{\bds\phi}> 0$, reproducing the
``dynamical Maxwell construction'' observed in ref. \cite{bdvhs}.  At
any rate, for the purposes of this work, above all it is important to
observe that, due to the quantum back--reaction, $M^2$ rapidly becomes
positive, within the so--called {\em spinodal time}
\cite{bdvhs,relax,devega2}, and then, for times before $\tau_L$, the
{\em weakly dissipative} regime takes place where $M^2$ oscillates
around zero with amplitude decreasing as $t^{-1}$ and a frequency
fixed by the largest spinodal wavevector, in complete agreement with
the infinite--volume results \cite{bdvhs}. Correspondingly, after
exponential grow until the spinodal time, the width of the zero--mode
grows on average linearly with time, reaching a maximum for $t
\simeq\tau_L$. Precisely, $\s_0$ performs small amplitude oscillations
with the same frequency of $M^2$ around a linear function of the form
$A + B t$, where $A, B \approx \lambda ^{-1/2}$ (see
Fig. \ref{fig:m0_l}), confirming what already found in
refs. \cite{bdvhs,relax}; then quite suddenly it turns down and enters
long irregular Poincar\'e--like cycles. Since the spinodal oscillation
frequency does not depend appreciably on $L$, the curves of $\s_0$ at
different values of $L$ are practically identical for
$t<\tau_L$. After a certain number of complete oscillations, a number
that scales linearly with $L$, a small change in the behavior of $M^2$
(see Fig. \ref{fig:usc_mass}) determines an inversion in $\s_0$ (see
Fig. \ref{fig:usc_zm}), evidently because of a phase crossover between
the two oscillation patterns. Shortly after $\tau_L$ dissipation
practically stops as the oscillations of $M^2$ stop decreasing in
amplitude and become more and more irregular, reflecting the same
irregularity in the evolution of the widths.

The main consequence of this numerical scenario is that the linear growth of the
zero--mode width at infinite volume cannot be consistently interpreted
as a form of Bose--Einstein Condensation (BEC) \cite{bdvhs}. If a
macroscopic condensation were really there, the zero mode would
develop a $\delta$ function in infinite volume, that would be
announced by a width of the zero mode growing to values $O(L^{3/2})$
at any given size $L$. Now, while it is surely true that when we push
$L$ to infinity, also the time $\tau_L$ tends to infinity, allowing
the zero mode to grow indefinitely, it is also true that, at any fixed
though arbitrarily large volume, the zero mode never reaches a width
$O(L^{3/2})$, just because $\tau_L \propto L$. In other words, if we
start from initial conditions where $\s_0$ is microscopic, then it
never becomes macroscopic later on.

\begin{figure}
\epsfig{file=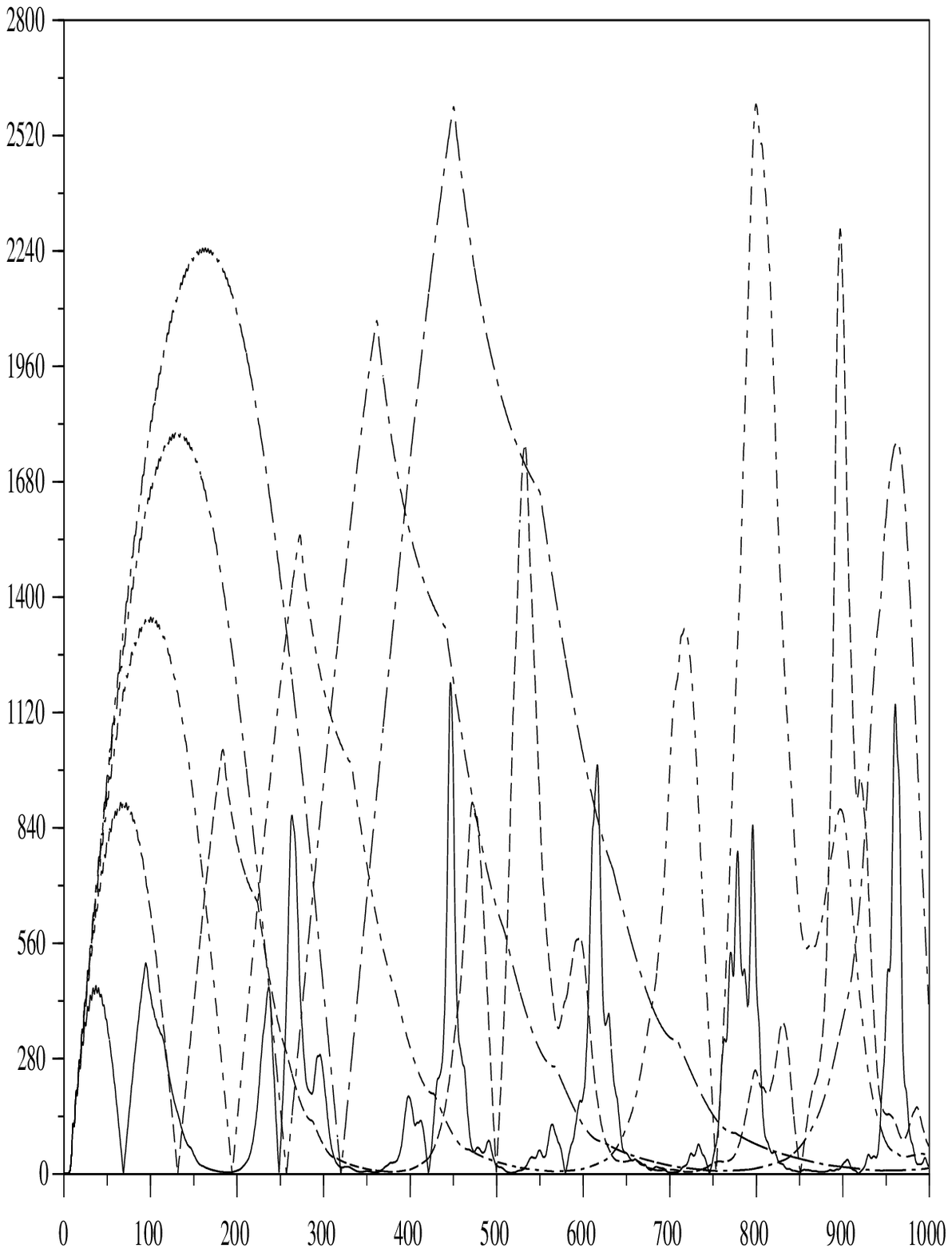,height=9cm,width=15cm}
\caption{\it Next--to--zero mode ($k=2\pi/L$) amplitude evolution for
different values of the size $L=20,40,60,80,100$, for $\lambda =
0.1$ and broken symmetry, with $\bar\phi=0$.}\label{fig:m1}
\end{figure}

\begin{figure}
\epsfig{file=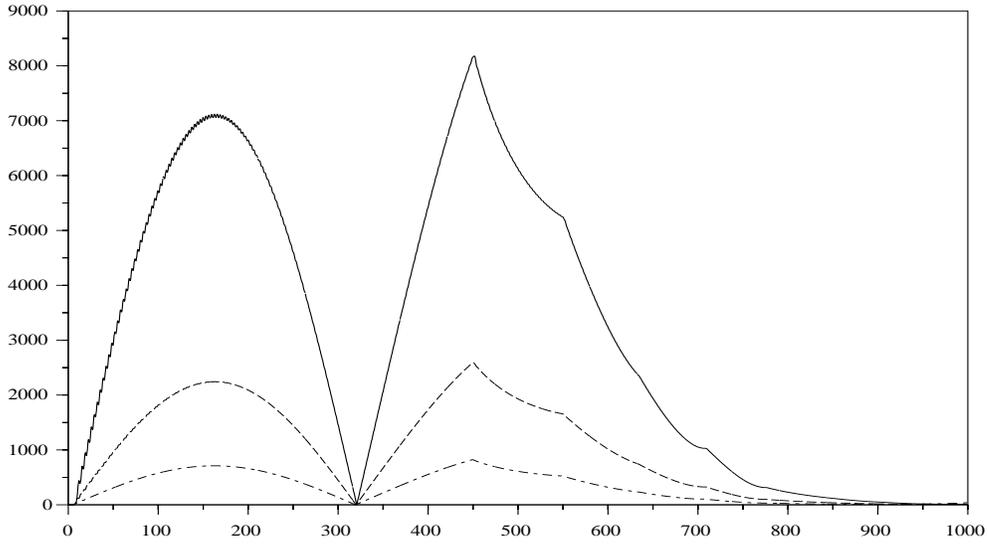,height=8.5cm,width=15cm}
\caption{\it Next--to--zero mode ($k=2\pi/L$) amplitude evolution for
different value of the renormalized coupling constant $\l=0.01,0.1,1$,
for $L=100$ and broken symmetry, with $\bar\phi=0$.}\label{fig:m1_l}
\end{figure}

On the other hand, looking at the behavior of the mode functions of
momenta $k=(2\pi/L)n$ for $n$ fixed but for different values of $L$,
one realizes that they obey a scaling similar to that observed for the
zero--mode: they oscillate in time with an amplitude and a period that
are $O(L)$ (see fig. \ref{fig:m1} and \ref{fig:m1_l}). Thus, each mode shows a
behavior that is exactly half a way between a macroscopic amplitude
[i.e. $O(L^{3/2})$] and a usual microscopic one [i.e. at most
$O(L^{1/2})$]. This means that the spectrum of the quantum
fluctuations at times of the order of the diverging volume can be
interpreted as a {\em massless} spectrum of {\em interacting}
Goldstone modes, because their power spectrum develops in the limit a
$1/k^2$ singularity, rather than the $1/k$ pole typical of free
massless modes. As a consequence the equal--time field correlation
function [see eq. (\ref{2pN})] will fall off as $|\bds x-\bds y|^{-1}$
for large separations smaller only than the diverging elapsed time.
This is in accord with what found in \cite{bdvhs}, where the same
conclusion where reached after a study of the correlation function for
the scalar field in infinite volume.

The fact that each mode never becomes macroscopic, if it started
microscopic, might be regarded as a manifestation of unitarity in the
large $N$ approximation: an initial gaussian state with only
microscopic widths satisfies clustering and clustering cannot be
spoiled by a unitary time evolution. As a consequence, in the
infinite--volume late--time dynamics, the zero--mode width $\s_0$ does
not play any special role and only the behavior of $\s_k$ as $k\to 0$
is relevant. As already stated above, it turns out from our numerics as
well as from refs. \cite{bdvhs,relax,tsunami} that this
behavior is of a novel type characteristic of the out--of--equilibrium
dynamics, with $\s_k \propto 1/k$.

\section{Time-dependent Hartree--Fock}\label{tdhf}

The main limitation of the large $N$ approximation, as far as the
evolution of the widths $\s_k$ is concerned, is in its intrinsic
gaussian nature. In fact, one might envisage a scenario in which,
while gaussian fluctuations stay microscopic, non--gaussian
fluctuations grow in time to a macroscopic size. In this section we 
therefore consider a time--dependent HF approximation capable in
principle of describing the dynamics of non--gaussian fluctuation of a
single scalar field with $\phi^4$ interaction.

As anticipated in section \ref{int}, we examine in this work only
states in which the scalar field has a
uniform, albeit possibly time--dependent expectation value.
In a tdHF approach we may then start from a wavefuction of the
factorized form (which would be exact for free fields)
\begin{equation}\label{wf}
	\Psi(\varphi)=\psi_0(\varphi_0)
		\prod_{k>0} \psi_k(\varphi_k,\varphi_{-k})
\end{equation}
The dependence of $\psi_k$ on its two arguments cannot be assumed to
factorize in general since space translations act as $SO(2)$ rotations
on $\varphi_k$ and $\varphi_{-k}$ (hence in case of translation
invariance $\psi_k$ depends only on $\varphi_k^2+\varphi_{-k}^2$).
The approximation consists in assuming this form as valid at all times
and imposing the stationarity condition on the action
\begin{equation}\label{varpri}
	\delta \int dt\, \vev{i\partial_t-H}=0 \;,\quad
	\vev{\cdot} \equiv \bra{\Psi(t)}\cdot\ket{\Psi(t)}
\end{equation}
with respect to variations of the functions $\psi_k$. To enforce a
uniform expectation value of $\phi$ we should add a Lagrange
multiplier term linear in the single modes expectations
$\vev{\varphi_k}$ for $k\neq 0$. The multiplier is then fixed at the
end to obtain $\vev{\varphi_k}=0$ for all $k\neq 0$. Actually one may
verify that this is equivalent to the simpler approach in which
$\vev{\varphi_k}$ is set to vanish for all $k\neq 0$ before any
variation. Then the only non trivial expectation value in the
Hamiltonian, namely that of the quartic term, assumes the form
\begin{equation}\label{vevphi4}
\begin{split}
	\int d^Dx\, \vev{\phi(x)^4} = & \frac1{L^D}
	\left[ \vev{\varphi_0^4}-3\vev{\varphi_0^2}^2 \right] +
	\frac3{2L^D} \sum_{k>0} \left[
	\vev{(\varphi_k^2+\varphi_{-k}^2)^2}-2\left(\vev{\varphi_k^2}
	+\vev{\varphi_{-k}^2}\right)^2 \right] \\ &+ 
	\frac3{L^D}\left(\sum_k\vev{\varphi_k^2}\right)^2 
\end{split}
\end{equation}
Notice that the terms in the first row would cancel completely out for
gaussian wavefunctions $\psi_k$ with zero mean value. The last term,
where the sum extends to all wavevectors $k$, corresponds instead to
the standard mean field replacement $\vev{\phi^4}\to
3\vev{\phi^2}^2$. The total energy of our trial state now reads
\begin{equation}\label{energy}
	E=\vev{H} =\dfrac12\sum_k \VEV{\pdif{^2}{\varphi_k^2}+
	(k^2+\mbare)\varphi_k^2} + \dfrac\lbare{4}\int d^Dx\, \vev{\phi(x)^4}
\end{equation}
and from the variational principle (\ref{varpri}) we obtain a set
of simple Schroedinger equations
\begin{equation}\label{Schroedinger}
	i\partial_t\psi_k = H_k \psi_k 
\end{equation}
\vskip -.4truecm
\begin{equation}\label{H_k}
\begin{split}
  H_0 &=-\dfrac12\pdif{^2}{\varphi_0^2}+\dfrac12 \om_0^2 \varphi_0^2
	+\dfrac{\lbare}{4L^D}\varphi_0^4   \\
  H_k &=-\dfrac12\left(\pdif{^2}{\varphi_k^2}+\pdif{^2}{\varphi_{-k}^2}\right)
	+\dfrac12 \om_k^2 (\varphi_k^2 + \varphi_{-k}^2)
	+\dfrac{3\lbare}{8L^D}\left(\varphi_k^2+\varphi_{-k}^2\right)^2 \\
\end{split}
\end{equation}
which are coupled in a mean--field way only through 
\begin{equation}\label{omvev}
	\om_k^2 = k^2+\mbare +3\lbare \Sigma_k  \;,\quad
	\Sigma_k= \dfrac1{L^D}\sum_{{q^2\le\Lambda^2}\atop
	{q\neq k,-k}}\vev{\varphi_q^2}
\end{equation}
and define the HF time evolution for the theory. By construction this
evolution conserves the total energy $E$ of eq. (\ref{energy}).

It should be stressed that in this particular tdHF approximation,
beside the mean--field back--reaction term $\Sigma_k$ of all other
modes on $\om_k^2$, we keep also the contribution of the {\em
diagonal} scattering through the diagonal quartic terms. In fact this
is why $\Sigma_k$ has no contribution from the $k-$mode itself: in a
gaussian approximation for the trial wavefunctions $\psi_k$ the
Hamiltonians $H_k$ would turn out to be harmonic, the quartic terms
being absent in favor of a complete back--reaction
\begin{equation}\label{Sigma}
	\Sigma = \Sigma_k + \dfrac{\vev{\varphi_k^2}+
	\vev{\varphi_{-k}^2}}{L^D} = \frac1{L^D}\sum_k\vev{\varphi_k^2}
\end{equation}
Of course the quartic self--interaction of the modes as well as the
difference between $\Sigma$ and $\Sigma_k$ are suppressed by a volume
effect and could be neglected in the infrared limit, provided all
wavefunctions $\psi_k$ stays concentrated on mode amplitudes
$\varphi_k$ of order smaller than $L^{D/2}$.  This is the
typical situation when all modes remain microscopic and the volume in
the denominators is compensated only through the summation over a
number of modes proportional to the volume itself, so that in the
limit $L\to\infty$ sums are replaced by integrals
\begin{equation*}
	\Sigma_k \to \Sigma \to \int_{k^2\le\Lambda^2} 
	\dfrac{d^Dk}{(2\pi)^D} \vev{\varphi_k^2}
\end{equation*}
Indeed we shall apply this picture to all modes with $k\neq 0$, while
we do expect exceptions for the zero--mode wavefunction $\psi_0$.

The treatment of ultraviolet divergences requires particular care,
since the HF approximation typically messes things up (see, for
instance, remarks in \cite{onhartree}). Following the same login of the
large $N$ approximation, we could take as renormalization condition
the requirement that the frequencies $\om_k^2$ are independent of
$\Lambda$, assuming that $\mbare$ and $\lbare$ are functions of
$\Lambda$ itself and of renormalized $\Lambda-$independent parameters
$m^2$ and $\l$ such that
\begin{equation}\label{renomvev}
	\om_k^2 = k^2 +m^2 +3\l\left[\Sigma_k\right]_{\text{finite}}
\end{equation}
where by $[.]_{\text{finite}}$ we mean the (scheme--dependent) finite
part of some possibly ultraviolet divergent quantity.  Unfortunately
this would not be enough to make the spectrum of energy differences
cutoff--independent, because of the bare coupling constant $\lbare$ in
front of the quartic terms in $H_k$ and the difference between
$\Sigma$ and $\Sigma_k$ [such problem does not exist in large $N$
because that is a purely gaussian approximation].  Again this would
not be a problem whenever these terms become negligible as
$L\to\infty$. At any rate, to be ready to handle the cases when this
is not actually true and to define an ultraviolet--finite model also
at finite volume, we shall by hand modify eq. (\ref{vevphi4}) as
follows:
\begin{equation*}
\begin{split}
	\lbare\int d^Dx\, \vev{\phi(x)^4} = & \l L^{-D}\left\{
	\vev{\varphi_0^4}-3\vev{\varphi_0^2}^2 + \tfrac32
	\sum\limits_{k>0} \left[
	\vev{(\varphi_k^2+\varphi_{-k}^2)^2}-2\left(\vev{\varphi_k^2}
	+\vev{\varphi_{-k}^2}\right)^2 \right]\right\} \\ &+ 
	3\lbare \,L^D\Sigma^2 
\end{split}
\end{equation*}
We keep the bare coupling constant in front of the term containing
$\Sigma^2$ because that part of the hamiltonian is properly
renormalized by means of the usual {\em cactus} resummation
\cite{cactus} which corresponds to the standard HF approximation. On
the other hand, within the same approximation, it is not possible to
renormalize the part in curly brackets of the equation above, because
of the factorized form (\ref{wf}) that we have assumed for the
wavefunction of the system. In fact, the $4-$legs vertices in the
curly brackets are diagonal in momentum space; when we contract two or
more vertices of this type, no sum over internal loop momenta is
produced, so that all higher order perturbation terms are suppressed
by volume effects. However, we know that in the complete theory, the
wavefunction is not factorized and loops contain all values of
momentum. This suggests that, in order to get a finite hamiltonian, we
need to introduce in the definition of our model some extra
resummation of Feynmann diagrams, that is not automatically contained
in this self--consistent HF approach. The only choice consistent with
the cactus resummation performed in the two--point function by the HF
scheme is the resummation of the $1$-loop {\em fish} diagram in the
four--point function. This amounts to the change from $\lbare $ to
$\l$ and it is enough to guarantee the ultraviolet finiteness of the
hamiltonian through the redefinition
\begin{equation}\label{H_k'}
	H_0 \to H_0+\dfrac{\l-\lbare}{4L^D}\varphi_0^4 \;,\quad
	H_k \to H_k +\dfrac{
	3(\l-\lbare)}{8L^D}\left(\varphi_k^2+\varphi_{-k}^2\right)^2
\end{equation}
At the same time the frequencies are now related to the widths
$\vev{\varphi_{-k}^2}$ by
\begin{equation}\label{omvev2}
\begin{split}
	\om_k^2 &= k^2+ M^2 - 3\l\, L^{-D}(\vev{\varphi_k^2}+
	\vev{\varphi_{-k}^2}) \;,\quad k>0 \\ M^2 &\equiv 
	\om_0^2 + 3\l\,L^{-D}\vev{\varphi_0^2}= \mbare +3\lbare \Sigma 
\end{split}
\end{equation}
Apart for $O(L^{-D})$ corrections, $M$ plays the role of
time--dependent mass for modes with $k\neq 0$, in the harmonic
approximation.
  
In this new setup the conserved energy reads
\begin{equation}\label{menergy}
	E =\sum_{k\ge 0}\vev{H_k} -\tfrac34 \lbare\,L^D\,\Sigma^2 +
	\tfrac34\l\,L^{-D}\left[\vev{\varphi_0^2}^2 + \sum\limits_{k>0}
	\left(\vev{\varphi_k^2}+\vev{\varphi_{-k}^2}\right)^2\right]
\end{equation}
Since the gap--like equations (\ref{omvev2}) are state--dependent, we
have to perform the renormalization first for some reference quantum
state, that is for some specific collection of wavefunctions $\psi_k$;
as soon as $\mbare$ and $\lbare$ are determined as functions
$\Lambda$, ultraviolet finiteness will hold for the entire class of
states with the same ultraviolet properties of the reference
state. Then an obvious consistency check for our HF approximation is
that this class is closed under time evolution.

Rather than a single state, we choose as reference the family of
gaussian states parametrized by the uniform expectation value
$\vev{\phi(x)}=L^{-D/2}\vev{\varphi_0}=\bar\phi$ (recall that we have
$\vev{\varphi_k}=0$ when $k\neq0$ by assumption) and such the HF
energy $E$ is as small as possible for fixed $\bar\phi$.  Then, apart
from a translation by $L^{D/2}\bar\phi$ on $\varphi_0$, these gaussian
$\psi_k$ are ground state eigenfunctions of the harmonic Hamiltonians
obtained from $H_k$ by dropping the quartic terms. Because of the
$k^2$ in the frequencies we expect these gaussian states to dominate
in the ultraviolet limit also at finite volume (as discussed above
they should dominate in the infinite--volume limit for any
$k\neq0$). Moreover, since now
\begin{equation}\label{gauss}
	\vev{\varphi_0^2} = L^D\bar\phi^2+ \dfrac1{2\om_0} \;,\quad 
	\vev{\varphi_{\pm k}^2} = \dfrac1{2\om_k} \;,\quad k\neq 0
\end{equation}
the relation (\ref{omvev2}) between frequencies and widths turn into the
single gap equation
\begin{equation}\label{gap}
	M^2 =\mbare+3\lbare\left(\bar\phi^2
	+\dfrac1{2L^D}\sum_{q^2\le\Lambda^2}\dfrac1{\sqrt{k^2+M^2}}\right)
\end{equation}
fixing the self-consistent value of $M$ as a function of $\bar\phi$.
It should be stressed that (\ref{omvev2}) turns through
eq. (\ref{gauss}) into the gap equation only because of the
requirement of energy minimization. Generic $\psi_k$, regarded as
initial conditions for the Schroedinger equations (\ref{Schroedinger}),
are in principle not subject to any gap equation.

The treatment now follows closely that in the large $N$ section, the
only difference being in the value of the coupling, now three times
larger.  In fact, in case of $O(N)$ symmetry, the quantum fluctuations
over a given background $\vev{\bds\phi(x)}=\bar{\bds\phi}$ decompose
for each $k$ into one longitudinal mode, parallel to $\bar{\bds\phi}$,
and $N-1$ transverse modes orthogonal to it; by boson combinatorics
the longitudinal mode couples to $\bar{\bds\phi}$ with strength
$3\lbare/N$ and decouple in the $N\to\infty$ limit, while the
transverse modes couple to $\bar{\bds\phi}$ with strength
$(N-1)\lbare/N\to \lbare$; when $N=1$ only the longitudinal mode is there. 

As $L\to\infty$, $\om^2_k \to k^2 +M^2$ and $M$ is exactly the
physical mass gap. Hence it must be $\Lambda-$independent.  At finite
$L$ we cannot use this request to determine $\mbare$ and $\lbare$,
since, unlike $M$, they cannot depend on the size $L$. At infinite volume
we obtain
\begin{equation}\label{M}
	M^2 =\mbare+3\lbare[\bar\phi^2+I_D(M^2,\Lambda)] \;,\quad
	I_D(z,\Lambda) \equiv \int_{k^2\le\Lambda^2} 
	\dfrac{d^Dk}{(2\pi)^D} \dfrac1{2\sqrt{k^2+z}}
\end{equation}
When $\bar\phi=0$ this equation fixes the bare mass to be
\begin{equation}\label{m2ren}
	\mbare= m^2 -3\lbare I_D(m^2,\Lambda)
\end{equation}
where $m=M(\bar\phi=0)$ may be identified with the equilibrium
physical mass of the scalar particles of the infinite--volume Fock
space without symmetry breaking (see below). Now, as in large $N$, the
coupling constant renormalization follows from the equalities
\begin{equation}\label{rengap}
\begin{split}
	M^2&=m^2+3\lbare[\bar\phi^2+I_D(M^2,\Lambda)-I_D(m^2,\Lambda)]\\
	&=m^2+3\l\,\bar\phi^2+3\l \left[ I_D(M^2,\Lambda)-
	I_D(m^2,\Lambda) \right]_{\text{finite}} 
\end{split}
\end{equation}
and reads when $D=3$
\begin{equation}\label{lren}
	\frac\l\lbare = 1-\frac{3\l}{8\pi^2} \log\frac{2\Lambda}{m\sqrt{e}} 
\end{equation}
that is the standard result of the one--loop renormalization group
\cite{zj}.  When $D=1$, that is a $1+1-$dimensional quantum field
theory, $I_D(M^2,\Lambda)-I_D(m^2,\Lambda)$ is already finite and the
dimensionfull couplings constant is not renormalized, $\lbare=\l$.

Again, the Landau pole in $\lbare$ prevents the actual UV limit
$\Lambda\to\infty$. Nonetheless, neglecting all inverse powers of the
UV cutoff when $D=3$, it is possible to rewrite the gap equation
(\ref{rengap}) as in eq. (\ref{nice}):
\begin{equation}\label{nice2}
	\frac{M^2}{\hat\l(M)} = \frac{m^2}{\hat\l(m)} + 3\,\bar\phi^2
\end{equation}
in terms of the one--loop running couplings constant
\begin{equation*}
	\hat\l(\mu) = \l \left[ 1 - \frac{3\l}{8\pi^2} 
	\log\frac{\mu}m \right]^{-1}
\end{equation*}
It is quite clear that the HF states for which the renormalization
just defined is sufficient are all those that are gaussian--dominated
in the ultraviolet, so that we have [compare to eq. (\ref{gauss})]
\begin{equation}\label{largek}
	\vev{\varphi_{\pm k}^2} \sim \dfrac1{2\om_k} 
	\;,\quad k^2 \sim \Lambda^2\;,\; \Lambda \to \infty
\end{equation}
If this property holds at a certain time, then it should hold at all
times, since the Schroedinger equations (\ref{Schroedinger}) are
indeed dominated by the quadratic term for large $\om_k$ and
$\om^2_k\sim k^2+\text{const}+O(k^{-1})$ as evident from
eq. (\ref{renomvev}). Thus this class of states is indeed closed under
time evolution and the parameterizations (\ref{m2ren}) and (\ref{lren})
make our tdHF approximation ultraviolet finite. Notice that the
requirement (\ref{largek}) effectively always imposes a gap equation
similar to eq. (\ref{gap}) in the deep ultraviolet.

Another simple check of the self--consistency of our approach,
including the change in selected places from $\lbare$ to $\l$, as
discussed above, follows from the energy calculation for the gaussian
states with $\vev{\phi(x)}=\bar\phi$ introduced above. Using
eq. (\ref{energy}) and the standard replacement of sums by integrals
in the infinite volume limit, we find
\begin{equation*}
	\E(\bar\phi) = \lim_{L\to\infty} \frac{E}{L ^D} = 
	\tfrac12\bar\phi^2(M^2-\l\bar\phi^2)+\tfrac12\int_{k^2\le\Lambda^2} 
	\dfrac{d^Dk}{(2\pi)^D} \,\sqrt{k^2+M^2} -\tfrac34 \lbare 
	\left[\bar\phi^2+I_D(M^2,\Lambda)\right]^2
\end{equation*}
where $M=M(\bar\phi)$ depends on $\bar\phi$ through the gap equation
(\ref{rengap}). The explicit calculation of the integrals involved
shows that the energy density difference $\E(\bar\phi)-\E(0)$ [which
for unbroken symmetry is nothing but the effective potential
$V_{\text{eff}}(\bar\phi)$], is indeed finite in the limit
$\Lambda\to\infty$, as required by a correct renormalization
scheme. Notice that the finiteness of the energy density difference
can be showed also by a simpler and more elegant argument, as
presented below in section \ref{ooed}. This check would fail instead when
$D=3$ if only the bare coupling constant $\lbare$ would appear in the
last formula.

The tdHF approximation derived above represents a huge simplification
with respect to the original problem, but its exact solution still
poses itself as a considerable challenge. As a matter of fact, a
numerical approach is perfectly possible within the capabilities of
modern computers, provided the number of equations
(\ref{Schroedinger}) is kept in the range of few thousands. As will
become clear later on, even this numerical workout will turn out not
to be really necessary in the form just alluded to, at least for the
purposes of this paper.

\subsection{On symmetry breaking}

Quite obviously, in a finite volume and with a UV cutoff there cannot
be any symmetry breaking, since the ground state is necessarily unique
and symmetric when the number of degrees of freedom is
finite \cite{gj}. However, we may handily envisage the situation which would
imply symmetry breaking when the volume diverges. 

Let us first consider the case that we would call of unbroken
symmetry. In this case the HF ground state is very close to the member
with $\bar\phi=0$ of the family of gaussian states introduced
before. The difference is entirely due to the quartic terms in $H_k$.
This correction vanish when $L\to\infty$, since all wavefunctions
$\psi_k$ have $L-$independent widths, so that one directly obtains the
symmetric vacuum state with all the right properties of the vacuum
(translation invariance, unicity, etc.)  upon which a standard scalar
massive particle Fock space can be based. The HF approximation then
turns out to be equivalent to the resummation of all ``cactus
diagrams'' for the particle self--energy \cite{cactus}. In a finite
volume, the crucial property of this symmetric vacuum is that all
frequencies $\om_k^2$ are strictly positive. The generalization to
non--equilibrium initial states with $\bar\phi\neq0$ is rather
trivial: it amounts to a shift by $L^{D/2}\bar\phi$ on
$\psi_0(\varphi_0)$. In the limit $L\to\infty$ we should express
$\psi_0$ as a function of $\xi=L^{-D/2}\varphi_0$ so that,
$|\psi_0(\xi)|^2\to \delta(\xi-\bar\phi)$, while all other
wavefunctions $\psi_k$ will reconstruct the gaussian wavefunctional
corresponding to the vacuum $\ket{0,M}$ of a free massive scalar
theory whose mass $M=M(\phi)$ solves the gap equation
(\ref{rengap}).  The absence of $\psi_0$ in $\ket{0,M}$ is
irrelevant in the infinite volume limit, since
$\vev{\varphi_0^2}=L^D\bar\phi^2 +$ terms of order $L^0$. The effective
potential $V_{\text{eff}}(\bar\phi)= \E(\bar\phi)-\E(0)$, where
$\E(\bar\phi)$ is the lowest energy density at fixed $\bar\phi$ and
infinite volume, is manifestly a convex function with a unique minimum
in $\bar\phi=0$.

Now let us consider a different situation in which one or more of the
$\om_k^2$ are negative. Quite evidently, this might happen only for
$k$ small enough, due to the $k^2$ in the gap equation [thus
eq. (\ref{largek}) remains valid and the ultraviolet renormalization
is the same as for unbroken symmetry]. Actually we assume here that
only $\om_0^2<0$, postponing the general analysis.  Now the quartic
term in $H_0$ cannot be neglected as $L\to\infty$, since in the ground
state $\psi_0$ is symmetrically concentrated around the two minima of
the potential $\frac12 \om_0^2 \varphi_0^2 +\frac\l{4L^D}\varphi_0^4$,
that is $\varphi_0=\pm(-\om_0^2L^D/\l)^{1/2}$.  If we scale
$\varphi_0$ as $\varphi_0=L^{D/2}\xi$ then $H_0$ becomes
\begin{equation}\label{Hscaled}
  H_0 = -\dfrac1{2L^D}\pdif{^2}{\xi^2}+\dfrac{L^D}2 \left(
	\om_0^2\, \xi^2 +\dfrac\l2\xi^4 \right)
\end{equation}
so that the larger $L$ grows the narrower $\psi_0(\xi)$ becomes around
the two minima $\xi=\pm(-\om_0^2/\l)^{1/2}$. In particular
$\vev{\xi^2}\to -\om_0^2/\l$ when $L\to\infty$ and
$\vev{\varphi_0^2}\simeq L^D\vev{\xi^2}$. Moreover, the energy gap
between the ground state of $H_0$ and its first, odd excited state as
well as difference between the relative probability distributions for $\xi$
vanish exponentially fast in the volume $L^D$. 
  
Since by hypothesis all $\om_k^2$ with $k\neq0$ are strictly positive,
the ground state $\psi_k$ with $k\neq0$ are asymptotically gaussian when
$L\to\infty$ and the relations (\ref{omvev2}) tend to the form 
\begin{equation*}
\begin{split}
	\om_k^2 &= k^2+ M^2  \equiv  k^2 +m^2 \\
	M^2 &= -2\om_0^2 = \mbare + 3\lbare (L^{-D}\vev{\varphi_0^2}
	+ \Sigma_0) = \mbare + 3\lbare\om_0^2+ 3\lbare I_D(m^2,\Lambda)] 
\end{split}
\end{equation*}  
This implies the identification $\om_0^2=-m^2/2$ and the bare mass
parameterization
\begin{equation}\label{brokenm}
	\mbare = \left(1-\tfrac32 \lbare/\l\right)m^2 -3\lbare I_D(m^2,\Lambda)
\end{equation}
characteristic of a negative $\om_0^2$ [compare to eq. (\ref{m2ren})],
with $m$ the physical equilibrium mass of the scalar particle, as in
the unbroken symmetry case. The coupling constant renormalization is
the same as in eq. (\ref{lren}) as may be verified by generalizing to
the minimum energy states with given field expectation value
$\bar\phi$; this minimum energy is nothing but the effective potential
$V_{\text{eff}}(\bar\phi)$; of course, since $\psi_0$ is no longer
asymptotically gaussian, we cannot simply shift it by $L^{D/2}\bar\phi$
but, due to the concentration of $\psi_0$ on classical minima as
$L\to\infty$, one readily finds that $V_{\text{eff}}(\bar\phi)$ is the
convex envelope of the classical potential, that is its Maxwell
construction. Hence we find
\begin{equation*}
	\vev{\varphi_0^2} \underset{L\to\infty}\sim
	\begin{cases}
	-L^D\om_0^2/\l \;,\; & \l\bar\phi^2\le-\om_0^2 \\
	L^D\bar\phi^2  \;,\; & \l\bar\phi^2>-\om_0^2
	\end{cases}
\end{equation*}
and the gap equation for the $\bar\phi-$dependent mass $M$ can be
written, in terms of the step function $\Theta$ and the extremal
ground state field expectation value $v=m/\sqrt{2\l}$,
\begin{equation}\label{gapbroken}
	M^2 = m^2 + 3\lbare(\bar\phi^2-v^2) \,
	\Theta(\bar\phi^2-v^2) +3\lbare \left[ 
	I_D(M^2,\Lambda)-I_D(m^2,\Lambda) \right] 
\end{equation}
We see that the specific bare mass parameterization (\ref{brokenm})
guarantees the non--renormalization of the tree--level relation
$v^2=m^2/2\l$ ensuing from the typical symmetry breaking classical
potential $V(\phi)=\frac14\l(\phi^2-v^2)^2$. With the same finite part
prescription as in eq. (\ref{rengap}), the gap equation
(\ref{gapbroken}) leads to the standard coupling constant
renormalization (\ref{lren}) when $D=3$.

In terms of the probability distributions $|\psi_0(\xi)|^2$ for the
scaled amplitude $\xi=L^{-D/2}\varphi_0$, the Maxwell construction
corresponds to the limiting form
\begin{equation}\label{limitform}
	|\psi_0(\xi)|^2 \underset{L\to\infty}\sim
	\begin{cases}
	\tfrac12(1+\bar\phi/v)\delta(\xi-v)+ \tfrac12(1-\bar\phi/v)\,
	\delta(\xi+v) \;,\; &\bar\phi^2 \le v^2 \\
	\delta(\xi-\bar\phi) \;,\;  &\bar\phi^2 > v^2 
	\end{cases}
\end{equation}
On the other hand, if $\om^2_0$ is indeed the
only negative squared frequency, the $k\ne0$ part of this minimum
energy state with arbitrary $\bar\phi=\vev{\phi(x)}$ is better and
better approximated as $\L\to\infty$ by the same gaussian state
$\ket{0,M}$ of the unbroken symmetry state. Only the effective mass
$M$ has a different dependence $M(\bar\phi)$, as given by the
gap equation (\ref{gapbroken}) proper of broken symmetry.

As in the large $N$ approach, at infinite volume we may write
\begin{equation*}
	\vev{\varphi_k^2} = C(\bar\phi)\,\delta^{(D)}(k)+
	\frac1{2\sqrt{k^2+M^2}}
\end{equation*}
where $C(\bar\phi)=\bar\phi^2$ in case of unbroken symmetry (that is
$\om^2_0>0$), while $C(\bar\phi)=\text{max}(v^2,\bar\phi^2)$
when $\om^2_0<0$.  This corresponds to the field correlation in space
\begin{equation*}
	\vev{\phi(x)\phi(y)}=\int\frac{d^Dk}{(2\pi)^D}
	\vev{\varphi_k^2} e^{ik\cdot(x-y)}= C(\bar\phi) +\Delta_D(x-y,M)
\end{equation*}
where $\Delta_D(x-y,M)$ is the massive free field equal--time two
points function in $D$ space dimensions, with self--consistent mass
$M$. As in large $N$, the requirement of clustering
\begin{equation*}
	\vev{\phi(x)\phi(y)} \to \vev{\phi(x)}^2 = v^2 
\end{equation*}
contradicts the infinite volume limit of 
\begin{equation*}
	\vev{\phi(x)}=L^{-D/2}\sum_k \vev{\phi_k}\,e^{ik\cdot x} =
	\vev{\varphi_0} = \bar\phi
\end{equation*}
except at the (now only two) extremal points $\bar\phi=\pm v$.
In fact we know that the $L\to\infty$ limit of the finite volume states with
$\bar\phi^2<v^2$ violate clustering, because the two peaks of
$\psi_0(\xi)$ have vanishing overlap in the limit and the first
excited state becomes degenerate with the vacuum: this implies that
the relative Hilbert space splits into two orthogonal Fock sectors
each exhibiting symmetry breaking, $\vev{\phi(x)}=\pm v$,
and corresponding to the two independent equal weight linear
combinations of the two degenerate vacuum states. The true vacuum is
either one of these symmetry broken states. Since the two Fock
sectors are not only orthogonal, but also superselected (no local
observable interpolates between them), linear combinations of any
pair of vectors from the two sectors are not distinguishable from
mixtures of states and clustering cannot hold in non--pure phases. It
is perhaps worth noticing also that the Maxwell construction for the
effective potential, in the infinite volume limit, is just a
straightforward manifestation of this fact and holds true, as such,
beyond the HF approximation.

To further clarify this point and in view of subsequent
applications, let us consider the probability distribution for the
smeared field $\phi_f=\int d^Dx\,\phi(x)f(x)$, where
\begin{equation*}
	f(x)=f(-x)=\frac1{L^D}\sum_k f_k \,e^{ik\cdot x} 
	\underset{L\to\infty}\sim \,\int \dfrac{d^Dk}{(2\pi)^D}
	\, \tilde f(k) \,e^{ik\cdot x} 
\end{equation*}
is a smooth real function with $\int d^Dx\,f(x)=1$ ({\em
i.e.} $f_0=1$) localized around the origin (which is good as any other
point owing to translation invariance). Neglecting in the infinite
volume limit the quartic corrections for all modes with $k\neq 0$, so
that the corresponding  ground state wavefunctions are
asymptotically gaussian, this probability distribution evaluates to
\begin{equation*}
	\text{Pr}(u\!<\!\phi_f\!<\!u+du)= \frac{du}{(2\pi\Sigma_f)^{1/2}} 
	\intf d\xi\, |\psi_0(\xi)|^2
	\exp\left\{\frac{-(u-\xi)^2}{2\Sigma_f} \right\}
\end{equation*}
where
\begin{equation*}
	\Sigma_f = \sum_{k\neq0}\vev{\varphi_k^2}\,f_k^2 
	\;\underset{L\to\infty}\sim \;\int \dfrac{d^Dk}{(2\pi)^D} 
	\,\dfrac{\tilde f(k)^2}{2\sqrt{k^2+m^2}}
\end{equation*}
In the unbroken symmetry case we have
$|\psi_0(\xi)|^2\sim\delta(\xi-\bar\phi)$ as $L\to\infty$, while
the limiting form (\ref{limitform}) holds for broken symmetry.
Thus we obtain
\begin{equation*}
	\text{Pr}(u\!<\!\phi_f\!<\!u+du) =p_f(u-\bar\phi)\,du \;,\quad 
	p_f(u) \equiv \left(2\pi\Sigma_f\right)^{-1/2}
	\exp\left(\frac{-u^2}{2\Sigma_f} \right)
\end{equation*}
for unbroken symmetry and 
\begin{equation*}
	\text{Pr}(u\!<\!\phi_f\!<\!u+du) =\begin{cases}
	\tfrac12(1+\bar\phi/v)\,p_f(u-v)\,du+\tfrac12(1-\bar\phi/v)
	\,p_f(u+v)\,du \;, &\bar\phi^2 \le v^2 \\
	p_f(u-\bar\phi)\,du\;, &\bar\phi^2 > v^2 
	\end{cases}
\end{equation*}
for broken symmetry. Notice that the momentum integration in the
expression for $\Sigma_f$ needs no longer an ultraviolet cutoff; of
course in the limit of delta--like test function $f(x)$, $\Sigma_f$
diverges and $p_f(u)$ flattens down to zero. The important observation
is that $\text{Pr}(u\!<\!\phi_f\!<\!u+du)$ has always a single peak
centered in $u=\bar\phi$ for unbroken symmetry, while for broken symmetry it
shows two peaks for $\bar\phi^2 \le v^2$ and  $\Sigma_f$ small enough.
For instance, if $\bar\phi=0$, then there are two peaks for 
$\Sigma_f<v^2$ [implying that $\tilde f(k)$ has a significant
support only up to wavevector $k$ of order $v$, when $D=3$, or
$m\exp(\text{const }v^2)$ when $D=1$].

To end the discussion on symmetry breaking, we may now verify the
validity of the assumption that only $\om_0^2$ is negative. In fact,
to any squared frequency $\om_k^2$ (with $k\neq0$) that stays strictly
negative as $L\to\infty$ there corresponds a wavefunction $\psi_k$
that concentrates on $\varphi_k^2+\varphi_{-k}^2=-\om_k^2L^D/\l$ ;
then eqs. (\ref{omvev2}) implies $-2\om_k^2 = k^2 + m^2$ for such
frequencies, while $\om_k^2 = k^2 +m^2$ for all frequencies with
positive squares; if there is a macroscopic number of negative
$\om_k^2$ (that is a number of order $L^D$), then the expression for
$\om_0^2$ in eq. (\ref{omvev2}) will contain a positive term of order
$L^D$ in the r.h.s., clearly incompatible with the requirements that
$\om_0^2<0$ and $\mbare$ be independent of $L$; if the number of
negative $\om_k^2$ is not macroscopic, then the largest wavevector
with a negative squared frequency tends to zero as $L\to\infty$ (the
negative $\om_k^2$ clearly pile in the infrared) and the situation is
equivalent, if not identical, to that discussed above with only
$\om_0^2<0$.

\subsection{Out--of--equilibrium dynamics}\label{ooed}

We considered above the lowest energy states with a predefinite
uniform field expectation value, $\vev{\phi(x)}=\bar\phi$, and
established how they drastically simplify in the infinite volume
limit.  For generic $\bar\phi$ these states are not stationary and
will evolve in time. By hypothesis $\psi_k$ is the ground state
eigenfunction of $H_k$ when $k>0$, and therefore $|\psi_k|^2$ would be
stationary for constant $\om_k$, but $\psi_0$ is not an eigenfunction
of $H_0$ unless $\bar\phi=0$. As soon as $|\psi_k|^2$ starts changing,
$\vev{\varphi_0^2}$ changes and so do all frequencies $\om_k$ which
are coupled to it by the eqs. (\ref{omvev2}). Thus the change
propagates to all wavefunctions.  The difficult task of studying this
dynamics can be simplified with the following scheme, that we might
call {\em gaussian approximation}. We first describe it and discuss
its validity later on.

Let us assume the usual gaussian form for the initial state [see
eq. (\ref{gauss}) and the discussion following it]. We know that it is
a good approximation to the lowest energy state with given
$\vev{\varphi_0}$ for unbroken symmetry, while it fails to be so for
broken symmetry, only as far as $\psi_0$ is concerned, unless
$\bar\phi^2 \geq v^2$. At any rate this is an acceptable initial state: the
question is about its time evolution. Suppose we adopt the harmonic
approximation for all $H_k$ with $k>0$ by dropping the quartic term.
This approximation will turn out to be valid only if the
width of $\psi_k$ do not grow up to the order $L^D$ (by symmetry the
center will stay in the origin). In practice we are now dealing with a
collection of harmonic oscillators with time--dependent frequencies and
the treatment is quite elementary: consider the simplest example of one quantum
degree of freedom described by the gaussian wavefunction
\begin{equation*}
	\psi(q,t)=\frac1{(2\pi\s^2)^{1/4}}
	\exp\left[-\frac12\left(\frac1{2\s^2}-i\frac{s}\s\right)q^2\right]
\end{equation*}
where $s$ and $\s$ are time--dependent. If the dynamics is determined
by the time--dependent harmonic hamiltonian
$\frac12[-\partial^2_q+\om(t)^2\,q^2]$, then the Schroedinger equation is
solved exactly provided that $s$ and $\s$ satisfy the classical
Hamilton equations
\begin{equation*}
	\dot\s = s \;,\quad \dot s = - \om^2\s + \frac1{4 \s^3}
\end{equation*}
It is not difficult to trace the ``centrifugal'' force $(4\s)^{-3}$
which prevent the vanishing of $\s$ to Heisenberg uncertainty
principle \cite{chkm,hkmp}.

The extension to our case with many degrees of freedom is
straightforward and we find the following system of equations
\begin{equation}\label{sk}
	i\pdif{}t \psi_0 = H_0\psi_0 \;,\quad
	\der{^2 \s_k}{t^2}= - \om_k^2\,\s_k + \frac1{4 \s_k^3} \;,\; k>0
\end{equation}
coupled in a mean--field way by the relations (\ref{omvev2}), which
now read
\begin{equation}\label{momvev2}
\begin{split}
	\om_k^2 &= k^2+ M^2 - 6\l\, L^{-D}\s_k^2 \;,\quad k>0 \\ 
	M^2 &= \mbare +3\lbare \left(L^{-D}\vev{\varphi_0^2} + \Sigma_0\right)
	\;,\quad \Sigma_0= \frac1{L^D} \sum_{k\ne 0}\s_k^2 
\end{split}
\end{equation}
This stage of a truly quantum zero--mode and classical modes with
$k>0$ does not appear fully consistent, since for large volumes some type of
classical or gaussian approximation should be considered for
$\varphi_0$ too. We may proceed in two (soon to be proven equivalent)
ways:
\begin{enumerate}
\item 
We shift $\varphi_0=L^{D/2}\bar\phi+\eta_0$ and then
deal with the quantum mode $\eta_0$ in the gaussian approximation,
taking into account that we must have $\vev{\eta_0}=0$ at all times.
This is most easily accomplished in the Heisenberg picture rather than
in the Schroedinger one adopted above. In any case we find that
the quantum dynamics of $\varphi_0$ is equivalent to the classical
dynamics of $\bar\phi$ and $\s_0\equiv\vev{\eta_0^2}^{1/2}$ described
by the ordinary differential equations
\begin{equation}\label{classical}
	\der{^2 \bar\phi}{t^2} = -\om_0^2\, \bar\phi -\l\, \bar\phi^3
	\;,\quad \der{^2 \s_0}{t^2}= - \om_0^2\,\s_0 + \frac1{4 \s_0^3}
\end{equation}
where $\om_0^2=M^2-3\l\,L^{-D}\vev{\varphi_0^2}$ and 
$\vev{\varphi_0^2} = L^D\bar\phi^2+ \s_0^2$.

\item 
We rescale $\varphi_0=L^{D/2}\xi$ right away, so that $H_0$ takes the
form of eq. (\ref{Hscaled}). Then $L\to\infty$ is the classical limit
such that $\psi_0(\xi)$ concentrates on $\xi=\bar\phi$ which evolves
according to the first of the classical equations in
(\ref{classical}). Since now there is no width associated to the
zero--mode, $\bar\phi$ is coupled only to the widths $\s_k$ with $k\neq 0$
by $\om_0^2=M^2-3\l\bar\phi^2$, while $M^2=\mbare
+3\lbare(\bar\phi^2+\Sigma_0)$.
\end{enumerate}

It is quite evident that these two approaches are completely
equivalent in the infinite volume limit, and both are good
approximation to the original tdHF Schroedinger equations, at least
provided that $\s_0^2$ stays  such that
$L^{-D}\s_0^2$ vanishes in the limit for any time. In this case we
have the evolution equations
\begin{equation}\label{Emotion}
	\der{^2 \bar\phi}{t^2} = (2\l\,\bar\phi^2 -M^2)\,\bar\phi \;,\quad
	\der{^2\s_k}{t^2} = - (k^2+M^2)\,\s_k + \frac1{4 \s_k^3} 	
\end{equation}
mean--field coupled by the $L\to\infty$ limit of eqs. (\ref{momvev2}),
namely
\begin{equation}\label{unbtdgap}
	M^2=m^2+3\lbare\left[\bar\phi^2 + \Sigma -I_D(m^2,\Lambda)\right]
\end{equation}
for unbroken symmetry [that is $\mbare$ as in eq. (\ref{m2ren})] or
\begin{equation}\label{btdgap}
	M^2=m^2+3\lbare\left[\bar\phi^2 -v^2 + 
	\Sigma -I_D(m^2,\Lambda)\right] \;,\quad m^2=2\l v^2
\end{equation}
for broken symmetry [that is $\mbare$ as in
eq. (\ref{brokenm})]. In any case we define
\begin{equation*}
	 \Sigma =\frac1{L^D} \sum_k\s_k^2 \;\underset{L\to\infty}\sim\;
	\int_{k^2\le\Lambda^2} \dfrac{d^Dk}{(2\pi)^D}\,\s_k^2
\end{equation*}
as the sum, or integral, over all microscopic gaussian widths
[N.B.:this definition differs from that given before in
eq. (\ref{Sigma}) by the classical term $\bar\phi^2$]. 

The conserved HF energy (density) corresponding to these equations of
motion reads
\begin{equation}\label{EHF}
\begin{split}
	\E &= \T +\V \;,\quad \T = \frac12(\dot{\bar{\phi}})^2 +
	\frac1{2L^D}\sum_k \dot\s_k^2  \\ \V &= \frac1{2L^D}\sum_k\left(
	 k^2\,\s_k^2 + \frac1{4\s^2_k}\right) +\tfrac12\mbare(\bar\phi^2+
	\Sigma) + \tfrac34\lbare (\bar\phi^2+\Sigma)^2 -\tfrac12\l\bar\phi^4
\end{split}
\end{equation}
Up to additive constants and terms vanishing in the infinite volume
limit, this expression agrees with the general HF energy of
eq. (\ref{menergy}) for gaussian wavefunctions.  It holds both for
unbroken and broken symmetry, the only difference being in the
parameterization of the bare mass in terms of UV cutoff and physical
mass, eqs. (\ref{m2ren}) and (\ref{brokenm}). The similarity to the
energy functional of the large $N$ approach, eq. (\ref{EHF}), is
evident; the only difference, apart from the obvious fact that
$\bar{\bds\phi}$ is a $O(n)$ vector rather than a single scalar, is in
the mean--field coupling $\s_k$--$\bar\phi$ and $\s_k$--$\Sigma$, due
to different coupling strength of transverse and longitudinal modes.

This difference between the HF approach for discrete symmetry ({\em
i.e} $N=1$) and the large $N$ method for the continuous
$O(N)$-symmetry is not very relevant if the symmetry is unbroken [it
does imply however a significantly slower dissipation to the modes of
the background energy density].  On the other hand it has a drastic
consequence on the equilibrium properties and on the
out--of--equilibrium dynamics in case of broken symmetry (see below),
since massless Goldstone bosons appear in the large $N$ approach,
while the HF treatment of the discrete symmetry case must exhibits a
mass also in the broken symmetry phase.

The analysis of physically viable initial conditions proceeds exactly
as in the large $N$ approach and will not be repeated here, except for
an important observation in case of broken symmetry. The formal energy
minimization w.r.t. $\s_k$ at fixed $\bar\phi$ leads again to
eqs. (\ref{inis}), and again these are acceptable initial conditions only if
the gap equation that follows from eq. (\ref{btdgap}) in the
$L\to\infty$ limit, namely
\begin{equation}\label{bgap}
	M^2 = m^2 + 3\lbare \left[\bar\phi^2-v^2 +
	I_D(M^2,\Lambda)-I_D(m^2,\Lambda) \right] 
\end{equation}
admits a nonnegative, physical solution for $M^2$. Notice that there
is no step function in eq. (\ref{bgap}), unlike the static case of
eq. (\ref{gapbroken}), because $\s_0^2$ was assumed to be microscopic,
so that the infinite volume $\s_k^2$ has no delta--like singularity in
$k=0$. Hence $M=m$ solves eq. (\ref{bgap}) only at the extremal points
$\bar\phi=\pm v$, while it was the solution of the static gap equation
(\ref{gapbroken}) throughout the Maxwell region $-v\le\bar\phi\le v$.
The important observation is that eq. (\ref{bgap}) admits a positive
solution for $M^2$ also within the Maxwell region. In fact it can be
written, neglecting as usual the inverse--power corrections in the UV
cutoff
\begin{equation}\label{bnice}
	\frac{M^2}{\hat\l(M)} = \frac{m^2}\l + 3\,(\bar\phi^2 -v^2) =
	3\,\bar\phi^2 -v^2
\end{equation}
and there exists indeed a positive solution $M^2$ smoothly connected
to the ground state, $\bar\phi^2=v^2$ and $M^2=m^2$, whenever
$\bar\phi^2\ge v^2/3$. The two intervals $v^2\ge\bar\phi^2\ge v^2/3$
correspond indeed to the metastability regions, while $\bar\phi^2<
v^2/3$ is the spinodal region, associated to a classical potential
proportional to $(\bar\phi^2 -v^2)^2$. This is another effect of the
different coupling of transverse and longitudinal modes: in the large
$N$ approach there are no metastability regions and the spinodal
region coincides with the Maxwell one. As in the large $N$ approach in
the spinodal interval there is no energy minimization possible, at
fixed background and for microscopic widths, so that a modified form
of the gap equation like eq. (\ref{newgap}) should be applied to
determine ultraviolet--finite initial conditions.

The main question now is: how will the gaussian widths $\s_k$ grow
with time, and in particular how will $\s_0$ grow in case of method 1
above, when we start from initial conditions where all widths are
microscopic? For the gaussian approximation to remain valid through
time, all $\s_k$, and in particular $\s_0$, must at least not become
macroscopic.  In fact we have already positively answered this
question in the large $N$ approach and the HF equations
(\ref{Emotion}) do not differ so much to expect the contrary now. In
particular, if we consider the special initial condition
$\bar\phi=\dot{\bar\phi}=0$, the dynamics of the widths is identical
to that in the large $N$ approach, apart from the rescaling by a
factor of three of the coupling constant.  Thus our HF approximation
confirms the large $N$ approach in the following sense: even if one
considers in the variational ansatz the possibility of non--gaussian
wavefunctionals, the time evolution from gaussian and microscopic
initial conditions is effectively restricted for large volumes to
non--macroscopic gaussians.

Strictly speaking, however, this might well not be enough, since the
infrared fluctuations do grow beyond the microscopic size to become of
order $L$. Then the quartic term in the low$-k$ Hamiltonians $H_k$ is
of order $L$ and therefore it is not negligible by itself in the
$L\to\infty$ limit, but only when compared to the quadratic term,
which {\em for a fixed $\om_k^2$ of order $1$} would be of order
$L^2$. But we know that, when $\bar\phi=0$, after the spinodal time and
before the $\tau_L$, the effective squared mass $M^2$ oscillates
around zero with amplitude decreasing as $t^{-1}$ and a frequency
fixed by the largest spinodal wavevector. In practice it is ``zero on
average'' and this reflect itself in the average linear growth of the
zero--mode fluctuations and, more generally, in the average harmonic
motion of the other widths with non--zero wavevectors. In particular
the modes with small wavevectors of order $L^{-1}$ feel an average
harmonic potential with $\om_k^2$ of order $L^{-2}$. This completely
compensate the amplitude of the mode itself, so that the quadratic
term in the low$-k$ Hamiltonians $H_k$ is of order $L^0$, much smaller
than the quartic term that was neglected beforehand in the gaussians
approximation. Clearly the approximation itself no longer appear fully
justified and a more delicate analysis is required. We intend to
return on this issue in a future work, restricting in the next section
our observations on some rather peculiar consequences of the gaussians
approximation that provide further evidence for its internal
inconsistency.

\subsection{Effective potential and late--time evolution}\label{ep}

By definition, the gaussian approximation of the effective potential
$V_{\text{eff}}(\bar\phi)$ coincides with the infinite--volume limit
of the potential energy $\V(\bar\phi,\{\s_k\})$ of eq. (\ref{EHF})
when the widths are of the $\bar\phi-$dependent, energy--minimizing
form (\ref{inis}) with the gap equation for $M^2$ admitting a
nonnegative solution. As we have seen, this holds true in the unbroken
symmetry case for any value of the background $\bar\phi$, so that the
gaussian $V_{\text{eff}}$ is identical to the HF one, since all
wavefunctions $\psi_k$ are asymptotically gaussians as
$L\to\infty$. In the presence of symmetry breaking instead, this
agreement holds true only for $\bar\phi^2\ge v^2$; for
$v^2/3\le\bar\phi^2<v^2$ the gaussian $V_{\text{eff}}$ exists but is
larger than the HF potential, which is already flat. In fact, for any 
$\bar\phi^2\ge v^2/3$, we may write the gaussian $V_{\text{eff}}$ as
\begin{equation*}
	V_{\text{eff}}(\bar\phi) = V_{\text{eff}}(-\bar\phi) =
	V_{\text{eff}}(v) + \int_v^{|\bar\phi|}\,du\,u[M(u)^2-2\l\,u^2] 
\end{equation*}
where $M(u)^2$ solves the gap equation (\ref{bnice}), namely
$M(u)^2=\hat\l(M(u))(3u^2-v^2)$. In each of the two disjoint regions
of definition this potential is smooth and convex, with unique minima
in $+v$ and $-v$, respectively. Its HF counterpart is identical for
$\bar\phi^2\ge v^2$, while it takes the constant value
$V_{\text{eff}}(v)$ throughout the internal region $\bar\phi^2<v^2$.
On the other hand the gaussian $V_{\text{eff}}$ cannot be defined
in the spinodal region $\bar\phi^2<v^2/3$, where the gap equation does
not admit a nonnegative solution in the physical region far away from
the Landau pole.

Let us first compare this HF situation with that of large $N$.  There
the different coupling of the transverse modes, three time smaller
than the HF longitudinal coupling, has two main consequences at the
static level: the gap equation (\ref{nice}) does not admit nonnegative
solutions for $\bar{\bds\phi}^2<v^2$, so that the spinodal region
coincides with the region in which the effective potential is flat,
and the physical mass vanishes. The out--of--equilibrium counterpart
of this is the dynamical Maxwell construction: when the initial
conditions are such that $\bar{\bds\phi}^2$ has a limit for
$t\to\infty$, all possible asymptotic values exactly span the flatness
region (and the effective mass vanishes in the limit). In practice
this means that $|\bar{\bds\phi}|$ is not the true dynamical order
parameter, whose large time limit coincides with $v$, the equilibrium
field expectation value in a pure phase. Rather, one should consider
as order parameter the renormalized local (squared) width
\begin{equation*}
	\lim_{N \to \infty} \frac{\vev{\bds{\phi}(x)
	\cdot\bds{\phi}(x)}_{\text{R}}}{N} = \bar{\bds\phi}^2 +
	\Sigma_{\text{R}} =  v^2 + \frac{M^2}\l
\end{equation*}
where the last equality follows from the definition itself of the
effective mass $M$ [see eq. (\ref{Nbgap})]. Since $M$ vanishes as
$t\to\infty$ when $\bar{\bds\phi}^2$ tends to a limit within the
flatness region, we find the renormalized local width tends to
the correct value $v$ which characterizes the broken symmetry phase,
that is the bottom of the classical potential.

\begin{figure}
\epsfig{file=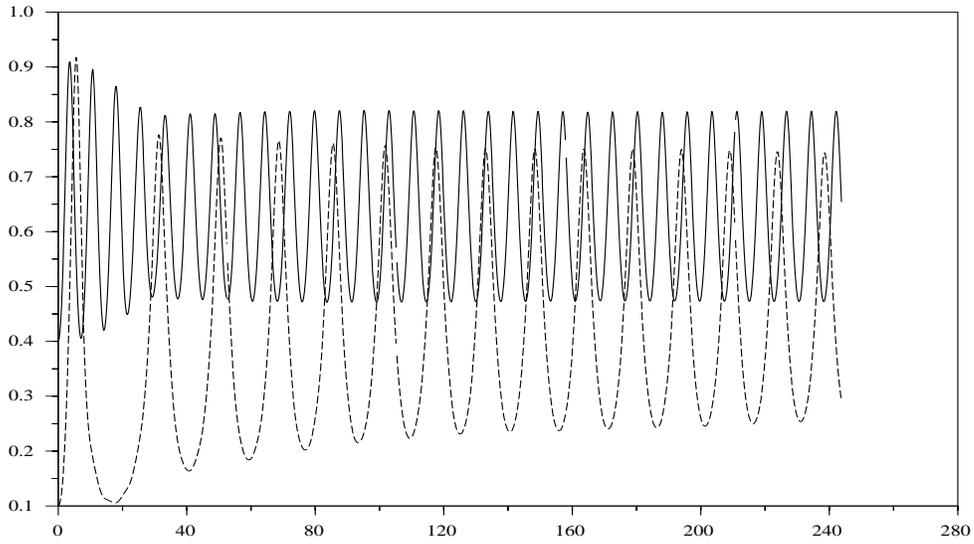,height=8.5cm,width=15cm}
\caption{\it Evolution of the background for two different initial
conditions within the spinodal interval, in the tdHF
approximation: $\bar\phi(t=0)=0.1$ (dotted line) and
$\bar\phi(t=0)=0.4$ (solid line). }\label{fig:max1}
\end{figure}

\begin{figure}
\epsfig{file=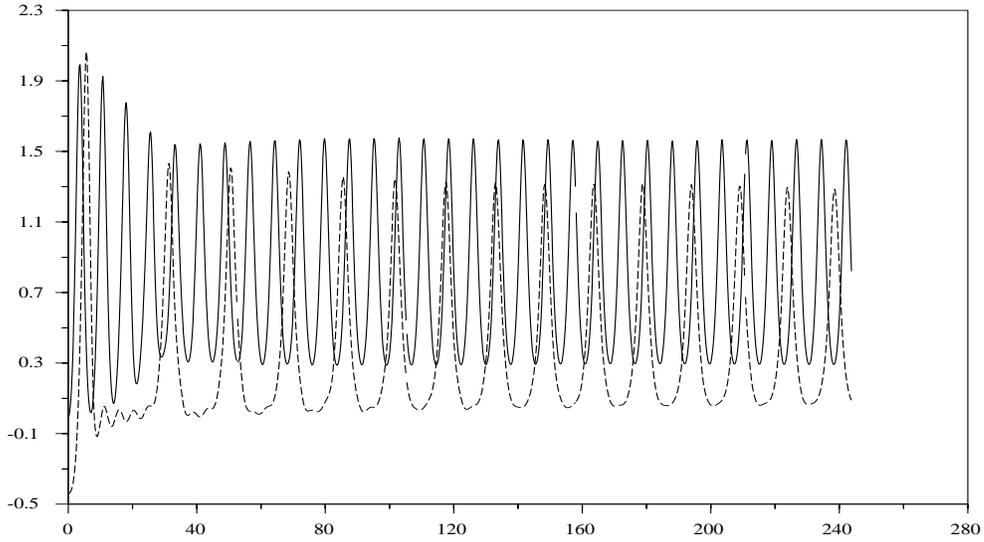,height=8.5cm,width=15cm}
\caption{\it Evolution of $M^2$ for the two initial
conditions of fig. \ref {fig:max1}. }\label{fig:max2}
\end{figure}

Let us now examine what happens instead in the dynamics of the HF
approximation, where at the static level the spinodal region
$\bar\phi^2<v^2/3$ is smaller than the flatness region
$\bar\phi^2<v^2$. Our (preliminary) numerical evidence shows that a
dynamical Maxwell construction take place as in the large $N$
approximation, but it covers only the spinodal region [see for
instance figs. \ref{fig:max1} and \ref{fig:max2}]. If the background $\bar{\phi}$
starts with zero velocity inside the spinodal interval, then it tends
to a limit within the same interval, the asymptotic force vanishing
because $M^2=2\l\,\bar{\phi}^2$ in the limit [see
eq. (\ref{Emotion})]. Hence we find that the order parameter
\begin{equation*}
	\vev{\phi(x)^2}_{\text{R}} = \bar{\phi}^2 +
	\Sigma_{\text{R}} = \frac{v^2}3 + 
	\frac{M^2-2\l\,\bar{\phi}^2}{3\l}  
\end{equation*}
[the relation (\ref{btdgap}) was used in the last equality] tends to
$v^2/3$ not $v^2$. It ``stops at the spinodal line''. This fact is at
the basis of the so--called {\em spinodal
inflation}\cite{holman}. Even without a conspicuous numerical evidence
for the full dynamical Maxwell construction, this results is
manifestly true when initially (and therefore at any time)
$\bar{\phi}=0=\dot{\bar{\phi}}$, because the equations of motion for
the widths are the same as in large $N$ and the effective mass only
differs by the factor of three in front of the quantum back--reaction
$\Sigma$. Since practically all the back--reaction takes place during
the exponential growth of the unstable spinodal modes, one could say
that, in the gaussian HF approximation, this back--reaction is to strong and
prevents the quantum fluctuations from sampling the true minima of the
classical potential. A side effect of this is that the effective mass
$M$ does not tend to its correct equilibrium value $m$ as
$t\to\infty$, unlike in the large $N$ approach.

In the previous section we have discussed the possible origin of the
problem: in our tdHF approach the initial gaussian wavefunctions are
allowed to evolve into non--gaussian forms, but they simply do not do
it in a macroscopic way, within a further harmonic approximation for
the evolution, so that in the infinite--volume limit they are
indistinguishable from gaussians at all times. But when $M^2$ is
on average not or order $L^0$, but much less, as it happens for
suitable initial conditions, infrared modes of order 
$L$ will be dominated by the quartic term in our Schroedinger
equations (\ref{Schroedinger}), showing a possible internal
inconsistency of the gaussians approximation.

Another interesting point concerns the dynamical Maxwell construction
itself, within the gaussians approximation. In fact it is not at all
trivial that the effective potential, in any of the approximation
previously discussed, does bear relevance on the asymptotic behavior
of the infinite--volume system whenever a fixed point is
approached. Strictly speaking in fact, even in such a special case the
effective potential has little to say on the dynamics, since it is
obtained from a static minimization and the energy is not at its
minimum at the initial time and is exactly conserved in the
evolution. On the other hand, if a solution of the equations of motion
(\ref{Emotion}) exists in which the background $\bar\phi$ tends to a
constant $\bar\phi_\infty$ as $t\to\infty$, one might expect that the
effective action (which however is nonlocal in time) somehow reduces
to a (infinite) multiple the effective potential, so that
$\bar\phi_\infty$ should be an extremal of the effective
potential. This is still an open question that deserve further analytic
studies and perhaps some further numerical confirmation.

\section{Conclusions and Perspectives}\label{conclusion}
In this work we have presented a rather detailed study of the
dynamical evolution out of equilibrium, in finite volume (a cubic box
of size $L$ in $3$D), for the $\phi^4$ QFT.  For comparison, we have
also analyzed some static characteristics of the theory both in
unbroken and broken symmetry phases. We have worked in two
non--perturbative approximation schemes, namely the large $N$
expansion in the leading order and a generalized time dependent
Hartree--Fock approximation.

We have reached two main conclusions.

The first, based on strong numerical evidence, is that the
linear growth of the zero--mode quantum fluctuations, observed already
in the large $N$ approach of refs. \cite{bdvhs,relax,tsunami}, cannot
be consistently interpreted as a new form of Bose--Einstein condensation. In
fact, in finite volume, $\s_0$ never grows to $O(L^{3/2})$ if it
starts from a microscopic value, that is at most of order
$L^{1/2}$. On the other hand all long--wavelength fluctuations rapidly
become of order $L$, signalling a novel infrared behavior quite
different from free massless fields at equilibrium [recall that the
large $N$ or HF approximations are of mean field type, with no direct
interaction among particle excitations]. This is in agreement with the
properties of the two--point function determined in \cite{bdvhs}.

The second point concerns the HF approach to the out--of--equilibrium
$\phi^4$ QFT. We have shown that, within a slightly enlarged tdHF
approach that allows for non--gaussian wavefunctions, one might
recover the usual gaussians HF approximation in a more controlled way,
realizing that the growth of long--wavelength fluctuations to order
$L$ in fact undermines the self--consistency of the gaussians
approximation itself. The first manifestation of this weakness is the 
curious ``stopping at the spinodal line'' of the width of the gaussian
quantum fluctuations. This does not happen in the large $N$ approach
because of different coupling of transverse mode (the only ones that
survive in the $N\to\infty$ limit) with respect to the longitudinal
modes of the $N=1$ case in the HF approach.

Clearly further study, both analytical and numerical, is needed in our
tdHF approach to better understand the dynamical evolution of quantum
fluctuations in the broken symmetry phase.  Another interesting
direction is the investigation, in our HF approximation, of the case
of finite $N$, in order to interpolate smoothly the results for $N=1$
to those of the $1/N$ approach. One first question concerns whether or
not the Goldstone theorem is respected in the HF approximation
\cite{onhartree}. It would be interesting also to study the dynamical
realization of the Goldstone paradigm, namely the asymptotic vanishing
of the effective mass in the broken symmetry phases, in different
models; this issue needs further study in the $2D$ case \cite{chkm},
where it is known that the Goldstone theorem is not valid.

Another open question concerns the connection between the minima of
the effective potential and the asymptotic values for the evolution of
the background, within the simplest gaussian approximation. As already
pointed out in \cite{bdvhs}, a dynamical Maxwell construction occurs
for the $O(N)$ model in infinite volume and at leading order in $1/N$
in case of broken symmetry, in the sense that any value of the
background within the spinodal region can be obtained as large time
limit of the evolution starting from suitable initial
conditions. Preliminary numerical evidence suggests that something
similar occurs also in the Hartree approximation for a single field,
but a more thorough and detailed analysis is needed.

\section{Acknowledgements}
C. D. thanks D. Boyanovsky, H. de Vega, R. Holman and M. Simionato
for very interesting discussions. C. D. and E. M. thank MURST and INFN for
financial support. Part of the work contained in this work has been
presented by E. M. at the 1999 CRM Summer School on ``Theoretical
Physics at the End of the XXth Century'', held in Banff (Alberta),
Canada, June 27 - July 10, 1999. E. M. thanks CRM (Universit\'e de
Montr\`eal) for partial financial support.

\appendix
\section{Details of the numerical analysis}\label{num}
We present here the precise form of the evolution equations for the
field background and the quantum mode widths, which control the
out--of--equilibrium dynamics of the $\phi ^4$ model in finite volume
at the leading order in the $1/N$ approach and in the tdHF approach,
as described in sections \ref{ooedN} and \ref{ooed}. We restrict here
our attention to the tridimensional case. 

Let us begin by noticing that each eigenvalue of the Laplacian operator in a
$3D$ finite volume is of the form $k_n ^2 = \left( \frac{2 \pi}{L}
\right) ^2 n$, where $n$ is a non--negative integer obtained as the sum of
three squared integers, $n = n _x ^2 + n _y ^2 + n _z ^2$. Then we
associate a degeneracy factor $g_n$ to
each eigenvalue, representing the
number of different ordered triples $(n _x, n _y, n _z)$ yielding the
same $n$. One may verify  that $g_n$ takes on the {\em continuum} value of
$4 \pi k ^2$ in the infinite volume limit, where $k=\left( \frac{2
\pi}{L} \right) ^2 n$ is kept fixed when $L \to \infty$.

Now, the system of coupled ordinary differential equations is, in case
of the large $N$ approach,
\begin{equation}\label{snumeq}
\left[ \frac{d ^2}{dt ^2} + M ^2\right] \phi
= 0 \;, \quad 
\left[ \frac{d ^2}{dt ^2} + \left( \frac{2 \pi}{L}
\right) ^2 n + M ^2 \right] \s_n - \frac1{4\s_n^3} =0
\end{equation}
where the index $n$ ranges from $0$ to ${\cal N}^2$, ${\cal N}=\Lambda
L/2\pi$ and $M^2(t)$ is defined by the eq. (\ref{Nunbgap}) in case of
unbroken symmetry and by eq. (\ref{Nbgap}) in case of broken symmetry.
The back--reaction $\Sigma$ reads, in the notations of this
appendix
\begin{equation*}
\Sigma = \frac{1}{L^D} \sum_{n=0}^{{\cal N}^2} g_n \s_n ^2 
\end{equation*}
Technically it is simpler to treat an equivalent set of
equations, which are formally linear and do not contain the singular
Heisenberg term $\propto \s_n^{-3}$. This is done by introducing the
complex mode amplitudes $z_n=\s_n\exp(i\t_n)$, where the phases $\t_n$
satisfy $\s_n^2\dot\t_n=1$. Then we find a discrete version of the
equations studied for instance in ref \cite{devega2},
\begin{equation}\label{numeq}
\left[ \frac{d ^2}{dt ^2} + \left( \frac{2 \pi}{L}
\right) ^2 n + M ^2 \right] z_n=0 \;,\quad 
\Sigma = \frac{1}{L^D} \sum_{n=0}^{{\cal N}^2} g_n |z_n| ^2
\end{equation}
subject to the Wronskian condition
\begin{equation*}
	z_n\,\dot{\bar{z_n}} - \bar{z_n}\,\dot z_n = -i
\end{equation*}
One realizes that the Heisenberg term in $\s_n$ corresponds to the 
centrifugal potential for the motion in the complex plane of $z_n$.

Let us now come back to the equations (\ref{numeq}). To solve these
evolution equations, we have to choose suitable initial conditions
respecting the Wronskian condition.  In case of unbroken symmetry,
once we have fixed the value of $\phi$ and its first time
derivative at initial time, the most natural way of fixing the initial
conditions for the $z_n$ is to require that they minimize the energy
at $t=0$. We can obviously fix the arbitrary phase in such a way to
have a real initial value for the complex mode functions
\begin{equation*}
z _n ( 0 ) = \frac{1}{\sqrt{2 \Omega _n}} \hspace{1 cm} \frac{d z _n}{dt} ( 0
) = \imath \sqrt{\frac{\Omega _n}{2}}
\end{equation*}
where $\Omega _n = \sqrt{k ^2 _n + M ^2 ( 0 )}$. The initial squared
effective mass $M ^2 (t= 0)$, has to be determined self-consistently,
by means of its definition (\ref{Nunbgap}).

In case of broken symmetry, the gap equation is a viable mean for
fixing the initial conditions only when $\phi$ lies outside the
spinodal region [cfr. eq (\ref{NiceN})]; otherwise, the gap equation
does not admit a positive solution for the squared effective mass. In
that case, we have to resort to other methods, in order to choose the
initial conditions. Following the discussion presented in \ref{ooedN},
one possible choice is to set $\s_k^2 = \frac1{2\sqrt{k^2+|M^2|}}$ for
$k^2<|M^2|$ in eq. (\ref{newgap}) and then solve the corresponding gap
equation (\ref{newgap}). An other acceptable choice would be to solve
the gap equation (\ref{newgap}), once we have set a massless spectrum
for all the spinodal modes but the zero mode, which is started from an
arbitrary, albeit microscopic, value.

There is actually a third possibility, that is in some sense half a
way between the unbroken and broken symmetry case. We could allow for
a time dependent bare mass, in such a way to simulate a sort of {\em
cooling down} of the system. In order to do that, we could start with
a unbroken symmetry bare potential (which fixes initial conditions
naturally via the gap equation) and then turn to a broken symmetry one
after a short interval of time. This evolution is achieved by a proper
interpolation in time of the two inequivalent parameterizations of the
bare mass, eqs. (\ref{Nm2ren}) and (\ref{minimum}).

We looked for the influence this different choices could produce in
the results and indeed they depend very little and only
quantitatively from the choice of initial condition we make.

As far as the numerical algorithm is concerned, we used a $4$th order
Runge-Kutta algorithm to solve the coupled differential equations
(\ref{numeq}), performing the computations in boxes of linear size
ranging from $L = 10$ to $L = 200$ and verifying the conservation of
the Wronskian to order $10^{-5}$. Typically, we have chosen values of
$\cal N$ corresponding to the UV cutoff $\Lambda$ equal to small
multiples of $m$ for unbroken symmetry and of $v\sqrt{\l}$ for broken
symmetry. In fact, the dynamics is very weakly sensitive to the
presence of the ultraviolet modes, once the proper subtractions are
performed. This is because only the modes inside the unstable
(forbidden or spinodal) band grow exponentially fast, reaching soon
non perturbative amplitudes (i.e. $ \approx \lambda ^{-1/2}$), while
the modes lying outside the unstable band remains perturbative,
contributing very little to the quantum back--reaction \cite{relax} and
weakly affecting the overall dynamics. The unique precaution to take
is that the initial conditions be such that the unstable band lay well
within the cutoff.

\vskip 0.5 truecm

\vskip 1 truecm

\vskip 1 truecm

\vskip 1 truecm

\end{document}